# Real-time Trajectory Optimization of Impaired Aircraft based on Steady State Manoeuvres

Ramin Norouzi,[1] Amirreza Kosari,[2]
*University of Tehran, Tehran, 1439957131, Iran*

Mark H. Lowenberg[3]
*University of Bristol, Bristol, England BS8 1TR, United Kingdom*

**Aircraft failures alter dynamics, diminishing manoeuvrability. Such manoeuvring flight envelope variations, governed by the aircraft's complex nonlinear dynamics, are unpredictable by pilots and existing flight management systems. To prevent in-flight Loss of Control, post-failure trajectories must be optimal, planned in real-time, avoid terrain, and adhere to the impaired aircraft's reduced manoeuvrability and dynamic constraints. This paper presents a novel real-time trajectory optimization method for impaired aircraft based on a combination of differential flatness theory, the pseudospectral method, nonlinear programming, and inverse dynamics. In the proposed method, which utilizes a high-fidelity nonlinear six degree-of-freedom model, to conform to aircraft's altered dynamics a sequence of trim points is selected from the impaired aircraft's manoeuvring flight envelope based on the chosen optimization criteria, ensuring that the resulting three-dimensional trajectory observes terrain avoidance. Then, the required control inputs are obtained for each manoeuvre in less than a second. The method is applied to the NASA Generic Transport Model with rudder failure near a complex mountainous terrain. Both an optimal one-piece trajectory and a near-optimal piecewise path consisting of several optimal trajectories, are generated in non-real-time and real-time, respectively, and compared. Results show that the near-optimal real-time trajectory retains up to 80% of the optimality.**

---

[1] Ph.D. Graduate, School of Aerospace Engineering; r.norouzi@ieee.org. Member AIAA.
[2] Associate Professor, School of Aerospace Engineering; kosari_a@ut.ac.ir.
[3] Professor of Flight Dynamics, School of Civil, Aerospace & Design Engineering; m.lowenberg@bristol.ac.uk. Senior Member AIAA.

## Nomenclature

| | | |
|---|---|---|
| $V$ | = | total airspeed, knot |
| $\alpha$ | = | angle of attack, deg |
| $\beta$ | = | sideslip angle, deg |
| $x, y, z$ | = | spatial position components in inertial coordinate system, m |
| $u, v, w$ | = | linear velocity components in body axes, m/s |
| $p, q, r$ | = | angular velocity components (roll rate, pitch rate, yaw rate, respectively) in body axes, deg/s |
| $\phi, \theta, \psi$ | = | Euler angles (roll, pitch, yaw, respectively), deg |
| $\gamma$ | = | flight path angle, deg |
| $\mathbb{x}$ | = | state vector |
| $\mathbb{u}$ | = | control vector |
| $\delta_e$ | = | elevator deflection angle, deg |
| $\delta_a$ | = | aileron deflection angle, deg |
| $\delta_r$ | = | rudder deflection angle, deg |
| $\delta_{th}$ | = | throttle setting (%) $\in [0, 1]$ |
| $\chi$ | = | heading angle, deg |
| $T$ | = | Thrust force, N |
| $L$ | = | Lift force, N |
| $D$ | = | Drag force, N |
| $h$ | = | flight altitude, ft |
| $h_{Terr}$ | = | terrain height, m |
| $h_{Cl}$ | = | safe clearance altitude, m |
| $m$ | = | aircraft mass, kg |
| $\bar{q}$ | = | dynamic pressure, kg/ms$^2$ |
| $\mathcal{X}$ | = | state vector approximation in orthogonal collocation |
| $\mathcal{U}$ | = | control vector approximation in orthogonal collocation |

# I. Introduction

As per the statistical data released by Boeing in September 2023, in-flight Loss of Control (LOC) -- accounting for 9 accidents and a total of 756 fatalities (approximately 28% of all fatal accidents) -- was the primary cause of fatal accidents involving commercial airliners between 2013 and 2022 [1]. Similarly, an Airbus report titled "A Statistical Analysis of Commercial Aviation Accidents 1958 - 2022," published in February 2023, identifies in-flight LOC as the leading category of fatal accidents spanning 2002 to 2022 [2]. Another study conducted by the UK Civil Aviation Authority in 2013, analysing fatal accidents from 2002 to 2013, reveals that nearly 40% of all fatal accidents were associated with loss of control, emphasizing its significance as the primary cause of these accidents [3]. Notably, despite an increase in flight frequency, there has been a downward trend in fatal accidents over this period, largely attributed to the emergence of more accurate flight control and safety systems, alongside more intelligent control automation systems [3]. Nevertheless, LOC remains the most significant contributor to the fatal accidents, despite all the enhancements in pilot training and aircraft systems. Generally, LOC occurs following an upset condition originating from technical failures such as control surface defects, or external events such as icing, or internal sources such as inappropriate pilot inputs, or from a combination of these factors [4].

In scenarios involving technical failures or external events, alterations occur in aircraft dynamics and parameters, leading to changes in the attainable flight envelope and its kinematic constraints. Typically, the impaired aircraft's degraded performance, governed by its altered nonlinear dynamics, is characterized by a more restricted flight envelope compared to the nominal one of an unimpaired aircraft [5]. As it is infeasible for the pilots to be trained for all possible failure situations, it is unlikely that the new aircraft dynamics and new flight envelope boundaries are accurately determined for pilots. Hence, under stress and excess workload, pilots try to swiftly plan a safe landing trajectory which may include an input or manoeuvre beyond the new admissible flight envelope of the impaired aircraft, thus leading to LOC [5]. Most systems designed so far in response to aircraft failures incorporate different adaptive or reconfigurable control methods. While these controllers are necessary for stabilizing the impaired aircraft and maintaining its controllability, they cannot guarantee that the sequence of manoeuvres (states) chosen by the pilot or the autopilot is feasible and that the planned post-failure trajectory is optimal based on the new altered dynamics of the impaired aircraft [6]. Consequently, handling an aircraft with degraded performance requires not only an adaptive/reconfigurable controller but also a system characterizing the reduced performance of the aircraft to augment the Flight Management System (FMS) [7]. Such a system should be also be capable of post-failure trajectory optimization based on the diminished manoeuvrability of the impaired aircraft.

Therefore, the challenge of preventing LOC-led-accidents through increasing the pilots' situational awareness and developing advanced FMS augmentation systems comprises two parts: 1) Developing advanced flight envelope protection systems which require robust and reliable post-failure flight envelope estimation capability. 2) Developing a real-time trajectory optimization approach that takes into account manoeuvrability limitations of impaired aircraft based on a high-fidelity nonlinear model, and real-time information of flight condition, type and degree of failure, and terrain. This is particularly challenging as the aircraft damages or failures can impose additional nonlinear influences on the stability and control from dynamic motions [8].

## A. Post-failure Manoeuvring Flight Envelope Estimation

As for the first part, various methods have been used in previous studies to evaluate the aircraft's flight envelope. A flight envelope is defined as a set of attainable trim states within a set of constraints. Loss of control may occur, once any of the constraints is violated [9]. Being attainable means the trim state falls within the region of attraction of the equilibrium state and the new stability margin is sufficient for the impaired aircraft to handle disturbances such as gusts [6]. Hence, one approach to aircraft flight envelop evaluation is determining the region of attraction of equilibrium points in the nonlinear system. Different methods have been proposed in literature for this purpose. For instance, in [10], subsets of the region of attraction were computed using real algebraic geometry theory by reformulating the problem as a Linear Matrix Inequality (LMI), whereas in [11, 12], the Lyapunov function method was utilized to estimate the attraction region of a stable equilibrium point in a nonlinear system. Additionally, in [13], a method was presented to assess the attraction region of equilibrium points of quadratic systems, requiring the solution of a feasibility problem involving LMI constraints. Also in [14], linear reachable set and nonlinear region of attraction techniques were employed to develop algorithms assessing the dynamic flight envelope of the NASA Generic Transport Model (GTM). In [15], means of sum of squares techniques were used to compute the inner estimates of the region of attraction. Recently, stable manifold theory was used to construct the region of attraction representing the dynamic flight envelope [16, 17].

Another approach is to evaluate the trim envelope by directly computing the achievable trim points based on high-fidelity models. For instance, in [18], steady states were calculated using Newton-Raphson method to evaluate the steady performance and maneuvering capabilities of an unimpaired aircraft in helical trajectories. Also in [19, 20], 3D manoeuvring flight envelopes of an impaired aircraft were evaluated by calculating all trim points in a point-by-point schema. Trim points are defined as steady state manoeuvres characterized by velocity, climb rate, turn rate and altitude, and derived by simultaneously minimizing the 6 degree-of-freedom (DoF) nonlinear equations of motion based on new dynamics of the damaged aircraft. In [7, 21, and 22], this method was applied to the NASA GTM with left wing damage to evaluate post-damage manoeuvring flight envelopes.

However, none of the aforementioned methods can be implemented onboard an impaired aircraft. This is due to their reliance on nonlinear models and extensive computational requirements, rendering them impractical for post-failure online flight envelope estimation. For instance online application of the reachability set method is limited to low-dimensional problems due to the "curse of dimensionality" [23] associated with the method [24].

To address the immediate need for flight envelope evaluation post-failure, researchers have endeavoured to develop computationally efficient flight envelope estimation techniques suitable for real-time implementation, integrating them into adaptive flight envelope protection systems. These classical techniques typically fall into one of three categories: local flight envelope estimation, employment of simplified models, and offline flight envelope databases.

The first category involves estimating the new flight envelope within the immediate vicinity of the current trim state. By confining the optimization space, online local flight envelope estimation becomes computationally feasible. This method entails progressively estimating local flight envelopes as new flight conditions are encountered. For example, in [25], the reachable set theory is employed to estimate local flight envelopes for airframe faults. Also, local flight envelope estimation has been utilized in a number of studies focused on developing adaptive flight planners [6, 26]. These studies employ linear discrete-time models to swiftly compute local manoeuvring flight envelopes comprising trim points defined as steady state manoeuvres characterized by velocity, climb rate, turn rate, and altitude. These evaluated trim points are then utilized as motion primitives by an adaptive flight planner to generate a feasible landing trajectory in post-failure motion planning.

The second category involves employing reduced complexity models instead of high-fidelity models, facilitating swift estimation of the trim envelope. For example, the reduced complexity model is obtained in [27] from the physical relationships between the variables deriving the slow dynamics of the aircraft state, wherein, thrust, angle of attack, sideslip angle, and bank angle are considered inputs, while the trim envelope is evaluated in terms of total airspeed and flight path angle. In [28], a point mass dynamic model is utilized, leveraging prior knowledge of the undamaged aircraft parameters to reduce the estimation problem's order. In this approach, a differential vortex lattice algorithm is employed for onboard flight envelope estimation of the impaired aircraft, with the estimated flight envelope represented in terms of altitude and airspeed once sideslip angle and bank angle are specified.

The third category involves generating an offline database of trim states for a priori known failures and deploying it onboard the aircraft. For instance, a method is proposed in [25] for online flight envelope interpolation following actuator fault detection, and in [29] for structural damages, where a database of flight envelopes associated with the most often occurring failure cases is created offline and accessed onboard the impaired aircraft to retrieve the closest envelopes to the occurred failure degree. Similarly, [30] introduces the Envelope-Aware Flight Management System (EA-FMS), an enhancement of conventional flight management systems that incorporates envelope estimation, flight planning, and safety assessment to mitigate loss of control incidents. This system assumes a priori characterization of failures, enabling offline envelope estimation across various failure degrees. The generated trim database is presumed applicable to any specific case online.

While the aforementioned methods offer numerical efficiency for onboard implementation, they have limited capability in real-time estimation of the entire flight envelope of the impaired aircraft based on high-fidelity models. Specifically, online local flight envelope estimation utilizing approaches outlined in [31, 32] enables fast evaluation of low-resolution flight envelopes in real-time, however evaluating high-resolution flight envelopes require spending considerable amount of time at each step that a local flight envelope is calculated, resulting in non-negligible accumulated time for the entire envelope estimation. Moreover, the selected final trim state could be infeasible as the entire flight envelope is not apparent initially. Therefore, progressing through the global flight envelope via step-by-step local flight envelope estimation may lead to surpassing the global flight envelope boundary eventually. Specifically, using local manoeuvring flight envelope estimation in the emergency path planning of the impaired aircraft could result in collision with terrain, as all manoeuvres within the local flight envelope at a future step may have paths intersecting with terrain despite the collision-free manoeuvres in the current local envelope.

Flight envelopes estimated using reduced complexity models are considerably simplified. Specifically, such envelopes, as presented in [33, 34] lack the steady state manoeuvre characteristics required for high fidelity nonlinear 6 degree-of-freedom post-failure emergency path planning.

Generating an offline database of impaired flight envelopes is applicable to failure scenarios characterized a priori and cannot accommodate unpredicted failures. Additionally, online interpolation of flight envelopes from an offline database necessitates carrying massive databases onboard with potentially inaccurate interpolation results if the offline flight envelopes are estimated in high resolution based on high-fidelity models. This is due to the fact that the boundaries of the impaired aircraft's global envelope shrink and drift within the steady-state-space, as discussed in our rigorous investigation of the unimpaired and impaired manoeuvring flight envelopes [35]. Moreover, since aircraft nonlinear dynamics directly influence the envelope variations with failure degrees, boundaries of high resolution manoeuvring flight envelopes at different failure degrees are not necessarily isotropically scaled relative to each other, as depicted in the database of the NASA GTM impaired manoeuvring flight envelopes and comparison plots therein [36]. Therefore, interpolating two flight envelopes with different sizes and locations in the steady-state-space, and different boundary shapes could result in a manoeuvring flight envelope with an inaccurate boundary.

An alternative approach to the aforementioned methods was proposed in one of our previous studies [37] enabling real-time estimation of high-fidelity global manoeuvring flight envelope (MFE) of impaired aircraft for any a priori unknown failure degree without the need of carrying massive databases onboard the aircraft. In this method, instead of evaluating the impaired aircraft's MFE through calculating all trim points in the MFE via trimming the impaired aircraft point-by-point at every steady state, artificial neural networks (ANNs) are utilized to directly estimate the boundary of the impaired aircraft's flight envelope.

In the proposed method, a large database of high-fidelity MFEs are computed offline for the unimpaired case and wide ranges of failure cases, specific number of points are distributed evenly on the boundaries of the 2D slices of the MFEs, and a system is designed containing the same number of neural networks where each network is assigned to a boundary point and trained offline utilizing the generated database to estimate in real-time the location of the corresponding boundary point at each specific a priori unknown failure onboard the aircraft. In the designed system, altitude and flight path angle along with other parameters which are determined according to the aircraft failure type and quantify the failure degree are considered as the inputs of each neural network. The location of each point in the $(V - \dot{\psi})$ 2D space is considered as the output of that point's network. As presented in [37], the proposed neural network-based flight envelope estimation method was capable of high accuracy computation of the high resolution global flight envelope in approximately 2.2 seconds, which, as detailed further in [37], is in compliance with the acceptable range of real time implementation of flight envelope estimation in related studies.

**B. Post-failure Trajectory Optimization**

In regards to the second part of the challenge of preventing LOC-led-accidents in commercial aircraft, various aircraft trajectory optimization approaches have been adopted or proposed in previous studies. Optimal path planning can be formulated as an optimal control problem. Trajectory optimization problems could have different applications based on the aircraft mission and defined cost function. One of the main applications of optimal path planning, which also considers terrain, is the Terrain Following and Terrain Avoidance (TF/TA) problem. Terrain Following (TF) is required in military missions particularly for low altitude flights aimed at avoiding radar detection, thereby necessitating Terrain Avoidance (TA). In such applications, aircraft trajectories are typically obtained as an approximation of the terrain profile demanding manoeuvres with complex dynamics. Tracking such intricate trajectories is generally infeasible for pilots, necessitating the need for a tracker to follow the control inputs. Time optimal terrain following problem was investigated in [38]. Effects of nonlinear engine dynamics were studied in [39] concluding that the conventional approach of neglecting engine dynamics in optimal path planning could result in inaccurate results. In [40], trajectory optimization problem in the presence of wind was solved using Shooting method. Avoiding airspace with non-convex constraints was studied in [41]. The terrain following problem was tackled in [42], integrating the terrain profile into aircraft kinematics. Further developments in [43] and [44] extended the approach to include moving targets and threats. TF control was resolved in [45] using a backstepping controller to track the angle of attack command. In [46], the TF problem was solved by parameterizing the aircraft altitude using third degree splines whose coefficients were determined by nonlinear programming (NLP). Steepest descent method was utilized in [47] to optimize the lateral motion of aircraft whereas parabolic flight segments were used for the longitudinal path and TA. Dynamics of a point-mass aircraft model were used in the optimization problem of [38], and trajectory tracking was accomplished in [48] via predictive control. In most of these studies, the performance index is chosen as a weighted combination of minimum time and minimum altitude.

In [49], a pseudospectral method combined with global polynomials was employed to derive constrained optimal trajectories taking into account the aircraft's control actuators. Direct transcription was utilized in [50] to obtain minimax optimal TF trajectories wherein terrain was represented by a cubic B-spline. For real-time TF/TA trajectory optimization, the Legendre pseudospectral method was employed on a 3-DoF model in [51], followed by inverse dynamics to determine the 6-DoF model parameters. Virtual Modified Terrain and a least squares technique were utilized in [52] to solve the optimal TF problem of a 2D point mass model. An optimal 3D TF/TA problem was investigated in [53] using an indirect approach.

Another important application of trajectory optimization, which has been less studied compared to the Terrain Following/Terrain Avoidance (TF/TA) problem, is emergency path planning for impaired aircraft. The constraints and requirements of this problem differ from the TF/TA problem because, in emergency path planning for impaired aircraft, the focus is on avoiding terrain rather than following it. However, due to intricacies imposed by inclusion of terrain into the emergency path planning problem, most studies have neglected the effect of terrain. For instance, in [54], a 3D Dubins path planning algorithm has been designed that generates no-thrust landing trajectories for a UAV to an appropriate landing site. Feasible 3D trajectories were obtained based on motion primitives and a graph search algorithm in [55]. Considered motion primitives or basic manoeuvres were: rectilinear path, glide path, and turning path with constant turn rate. In [56], an autonomous emergency motion planning was proposed for fixed-wing unmanned aircraft suffering loss-of-thrust. An optimal turn-back manoeuvre after engine failure was addressed in [57] via optimal control. In [58], Gauss Pseudospectral Method (GPM) was employed for optimal reconfiguration of spacecraft formations. In [59], the optimal emergency landing trajectory for an engine-out DC-9 aircraft was solved using GPM, though no terrain was taken into account. The study employed a traditional two-step approach to tackle the resulting NLP problem: first, the problem was solved with a limited number of nodes, and then the solution was used as the initial guess for solving the problem with a larger number of nodes.

The most relevant studies to this research are those focusing on emergency trajectory optimization for impaired commercial airliners. A method to determine safe parameter envelopes for emergency trajectories following a total engine flame out event was proposed in [60] assuming the envelope boundary to be a Pareto frontier of a multi-objective optimal control problem. Since the envelope boundary is estimated in approximately 45 minutes, it should be estimated offline and only serve as a dataset for the analysis of emergency landing scenarios. Yet the calculation of the landing trajectories can be performed in short amount of time. A reduced complexity aircraft model was used in this research, and terrain was not considered. In [61], 6-DoF flight dynamics was used to calculate and build a database of gliding footprints for an aircraft with thrust failure. An ANN model was trained with the footprint database enabling the prediction of the gliding footprint under the entire operational altitude and full radial direction while satisfying the real-time requirement. However, terrain was not taken into account. In [20], manoeuvres were chosen from the trim points of a reduced database of manoeuvring flight envelopes and their execution times were selected aiming to minimize the total flight time. To do so, kinematic equations were utilized to calculate the aircraft's location at the end of each trim manoeuvre. Even with the reduced database, the case studies required path planning times ranging from 20 to 60 minutes, making in-flight implementation impractical. In [62], aircraft experiencing total loss of thrust were studied, where trim points were chosen to form the best glide and Dubins curve trajectories toward the intended landing site. This research focused on geometrical path planning. While Dubins curves create smooth trajectories, they do not offer terrain avoidance due to their geometric nature. Dubins curves were also explored in [63-65] for emergency path planning. These studies focused on aircraft that could not maintain rectilinear flight due to the experienced failure, generating a geometric (kinematic) landing trajectory in real-time using the Turning Dubins Vehicle solver. This solver generates landing trajectories with minimum length.

A generalized transport aircraft with left wing structural damage was considered in [22, 7]. Specifically, in [22], assuming the impaired manoeuvring flight envelopes are known beforehand, a heuristic cost function was established based on several criteria, including altitude from the airport, heading difference with touchdown heading, distance to terrain, deviation from a desirable glide slope, and proximity to the flight envelope boundary. Trim states were explored and chosen as motion primitives forming the emergency landing path according to the cost function criteria. To enable real-time trajectory planning, the study relied on three main assumptions: first, that the manoeuvring flight envelope for the specific incurred failure was pre-calculated and available onboard the aircraft; second, that aircraft position displacement for each manoeuvre was pre-computed and stored in a database; and third, instead of examining the entire trim space, which would be impractical in real-time, an Artificial Potential Field algorithm was employed such that only the 27 trim manoeuvres surrounding each trim point (as per concentric cubes definition in [20]) were searched, the optimal manoeuvre was selected, the aircraft performed the selected manoeuvre, and the process was repeated until the aircraft reached the runway. The authors dismissed other search algorithms, such as A* and Dijkstra, due to their inability to conduct real-time searches within manoeuvring flight envelopes. It is unfeasible to predict all potential failure scenarios and evaluate their associated manoeuvring flight envelopes and store them in a database, ahead of time. This makes the first assumption impractical in real-world contexts. Additionally, the second assumption is flawed because the physical space trajectory should dictate the necessary manoeuvres, rather than being driven by them. As for the third assumption, it restricts true terrain avoidance since minimizing the terrain avoidance cost function is only done in the manoeuvres immediately around the current trim point, rather than across the entire search space. As noted in [26], a similar assumption to the third assumption in [22] can lead to a dead-end trajectory. This is due to the limited search instead of searching among all available trim manoeuvres —choosing a seemingly optimal manoeuvre at a particular point might initially appear to be the best for terrain avoidance, yet it could lead to other manoeuvres (not explored previously due to the adopted real-time approach) further along the path that all result in collision with terrain.

A second approach to terrain avoidance was presented in [26], where, after a trim manoeuvre is chosen at each step, a post-processing step is executed in physical space to verify that the chosen manoeuvre does not result in collision with terrain. If a collision is detected, an escape trajectory is executed. This method ensures that, if selecting a particular trim manoeuvre leads to a dead-end point several steps ahead where all surrounding manoeuvres would result in collision with terrain, an escape manoeuvre can be carried out to avoid it. As stated in [26], while this approach does not guarantee even locally optimal trajectories, it does at least prevent terrain collisions. Nonetheless, even the second approach cannot appropriately address the terrain avoidance problem and only offers a temporary fix which could result in accidents in emergency cases (e.g. minimum time or minimum fuel landing).

In this paper, the Radau Pseudospectral Method (RPM), known for its high accuracy and rapid convergence as one of the direct methods, is employed for real-time emergency landing path planning of an impaired commercial aircraft. The resulting trajectory consists of manoeuvres strictly within the impaired aircraft's manoeuvring flight envelope boundaries while adhering to terrain avoidance constraints. This proposed real-time trajectory optimization approach, combined with our previously developed real-time neural network-based manoeuvring flight envelope estimation method, forms an integrated solution to the LOC-led accidents prevention challenge. This integrated approach offers robust and reliable post-failure flight envelope estimation, alongside real-time trajectory optimization, accounting for the manoeuvrability limitations of impaired aircraft. It is worth mentioning that the trajectory optimization approach proposed in this paper is not exclusively reliant on our previously proposed real-time flight envelope estimation method and can also be enabled by other approaches that provide post-failure information on impaired aircraft's manoeuvrability, such as carrying an offline-generated database of manoeuvring flight envelopes onboard the aircraft.

The rest of this paper is organized as follows: In Section II, the proposed methodology is detailed. Section III discusses a case study involving the application of the proposed method to an impaired GTM with jammed rudder. Finally, Section IV concludes the paper.

## II.    Methodology

### A.  Manoeuvring Flight Envelope

As will be discussed in the following subsections, the proposed trajectory optimization approach of this study employs impaired aircraft's manoeuvring flight envelopes. Therefore, this subsection briefly describes the process in which manoeuvring flight envelopes are estimated. More details on this can be found in [6, 20, 35, and 66].

Manoeuvring flight envelopes (MFEs) are defined as boundaries containing steady state manoeuvres. Such steady state manoeuvres are referred to as trim points [5]. A steady state manoeuvre refers to the condition where each individual force and moment is either zero or constant while the net acting force and moment on the aircraft are zero. This requires rates of all linear and angular velocities as well as aerodynamic angles rates to be zero, whilst controls are fixed [67]. Thus, in the wind-axes coordinate system:

$$(\dot{p}, \dot{q}, \dot{r}) \equiv (\dot{V}, \dot{\alpha}, \dot{\beta}) = 0 \qquad (1)$$

(1), is a general definition of trim state. Therefore, to define the intended level/climbing/descending rectilinear and turning manoeuvres, additional constraints are considered as following:

$$\text{Level rectilinear: } \dot{\phi}, \dot{\theta}, \dot{\psi}, \gamma = 0 \tag{2}$$

$$\text{Climbing/descending rectilinear: } \dot{\phi}, \dot{\theta}, \dot{\psi} = 0, \ \gamma = cte \tag{3}$$

$$\text{Level turn: } \dot{\phi}, \dot{\theta}, \gamma = 0, \ \dot{\psi} = cte \tag{4}$$

$$\text{Climbing/descending turn: } \dot{\phi}, \dot{\theta} = 0, \ \dot{\psi}, \gamma = cte \tag{5}$$

where $cte$ denotes the constant value determined based on the specific steady state manoeuvre characteristics (i.e. climb rate and turn rate). The trim state definition in terms of the nonlinear aircraft equations of motion can be written as:

$$\dot{\mathbb{x}}_{trim} = f(\mathbb{x}_{trim}, \mathbb{u}_{trim}) = 0 \tag{6}$$

$$\mathbb{x}_{trim} = [V, \alpha, \beta, p, q, r, \phi, \theta]^T \tag{7}$$

$$\mathbb{u}_{trim} = [\delta_{th}, \delta_e, \delta_a, \delta_r]^T \tag{8}$$

$$(\gamma = \gamma^*), \ \ (\dot{\psi} = \dot{\psi}^*), \ \ (\dot{\phi}, \ \dot{\theta} = 0) \tag{9}$$

As the inertial position variables and yaw angle $(x, y, h, \psi)$ only specify the spatial orientation of the aircraft in 3D space, they have been omitted from the 12-dimensioal state vector in the context of trimmed flight, resulting in the reduced form in (7). $\gamma^*, \dot{\psi}^*$ in (9) are the desired constant values defining the steady state manoeuvres presented in (2)−(5). MFEs are comprised of trim states characterized by four parameters $(h^*, V^*, \gamma^*, \dot{\psi}^*)$. Therefore, these flight envelopes can be depicted as three-dimensional volumes at each constant flight altitude $h^*$. Since each trim condition must satisfy the aircraft equations of motion in (6), for each desired steady state manoeuvre, trim vectors $(\mathbb{x}_{trim}, \mathbb{u}_{trim})$ are determined by solving all aircraft nonlinear equations of motion $(\dot{\mathbb{x}}_{trim} = 0)$ simultaneously for the desired flight path angle, turn rate, and total airspeed $(\gamma^*, \dot{\psi}^*, V^*)$ at a specific altitude $h^*$.

Due to the nonlinearities of the equations of motion and their complexity it is not analytically possible to solve $(\dot{\mathbb{x}}_{trim} = 0)$, hence the feasible trim points are derived by numerically solving the corresponding constrained nonlinear optimization problem in which the cost function is defined as [7, 20, and 67]:

$$\Re(x, u) = \frac{1}{2} \dot{\mathbb{x}}_{trim}^T \mathcal{G} \dot{\mathbb{x}}_{trim} \tag{10}$$

where $\mathcal{G}$ is a positive definite weighting matrix specifying the state derivatives' contributions to the cost function $\Re$. Cost function $\Re$ is subject to the set of the equality constraints (11−13), the inequality constraints which are dictated by the physical limits on the control inputs (14), and the inequality constraints on the bank angle, angle of attack and the flight path angle (15−17).

(11) constrains the flying altitude, total airspeed, flight path angle, and turn rate to the intended trim state values denoted by the asterisk. (12) is obtained once the rate-of-climb (i.e. $V \sin \gamma$) constraint is solved for $\theta$ [67]. In case of a wings-level, non-sideslipping flight (12) reduces to $\theta = \alpha + \gamma$. The three equations in (13) are derived from the rotational kinematics by applying the steady-state manoeuvre constraint of $\dot{\phi} = \dot{\theta} = 0$. Control surface deflection limits in (14) are all in degree and have been set according to the corresponding values in the software model of the NASA GTM aircraft, *GTM-DesignSim*. Also throttle is in the range of [0, 1] (i.e. normalized to its maximum).

| | |
|---|---|
| $h - h^* = 0$ , $V - V^* = 0$ , $\gamma - \gamma^* = 0$ , $\dot{\psi} - \dot{\psi}^* = 0$ | (11) |
| $tan\theta - \dfrac{ab + sin\gamma^* \sqrt{a^2 - sin^2\gamma^* + b^2}}{a^2 - sin^2\gamma^*} = 0,\ \ \theta \neq \pm\pi/2$  <br><br> where, $\quad a = cos\alpha cos\beta,\ \ b = sin\phi sin\beta + cos\phi sin\alpha cos\beta$ | (12) |
| $p + \dot{\psi}^* sin\theta = 0$  <br><br> $q - \dot{\psi}^* cos\theta sin\phi = 0$  <br><br> $r - \dot{\psi}^* cos\theta cos\phi = 0$ | (13) |
| $\|\delta_{th} - 0.5\| \leq 0.5 \quad \|\delta_e\| \leq 30$  <br> $\|\delta_a\| \leq 20 \quad\quad\ \|\delta_r\| \leq 30$ | (14) |
| $-30° \leq \phi \leq 30°$ | (15) |
| $\alpha \leq 10.5°$ | (16) |
| $-5° \leq \gamma \leq 5°$ | (17) |

To evaluate the feasibility of each intended trim point, the constrained optimization problem defined by $\Re$ and the set of the equality and inequality constraints (11)-(17) is solved via a sequential quadratic programming (SQP) technique with a convergence criterion of $10^{-7}$. A trim state cannot be included within the boundaries of the MFE solely for being feasible. Feasibility is the necessary condition whilst stability is the sufficient condition for inclusion in the flight envelope. Stable trim points are generally preferred as the aircraft naturally tends to damp the effect of minor disturbances around them, while at unstable trim points the aircraft deviates from the trim state. Nevertheless, if the feasible and unstable trim state is controllable, i.e. if the linear perturbation system about this trim state has a full rank controllability matrix, it is accepted as part of the flight envelope. A nonlinear system is considered stable at a specific trim point, if the system inherently converges to the trim state when being in the vicinity of the trim point [68]. Therefore, to evaluate the stability of the aircraft at each feasible trim point $\mathbf{x}^*$, aircraft motion is approximated by linearizing the equations of motion about $\mathbf{x}^*$ via a linear perturbation method. Derived system eigenvalues determine if the considered trim point is stable. A trim state is stable if the state matrix has no positive real eigenvalues or complex eigenvalues with positive real part. If the investigated trim point $\mathbf{x}^*$ is found to be unstable then the controllability matrix of the perturbation system about $\mathbf{x}^*$ is formed and its rank is checked. In case of a full rank matrix, $\mathbf{x}^*$ is controllable and is included in the MFE, otherwise it will be excluded.

To evaluate 3D manoeuvring flight envelopes, the numerical procedure explained above is implemented iteratively for each triplet $(V^*, \gamma^*, \dot{\psi}^*)$ within the considered ranges. In this process, the successful optimization solution for the previous triplet is considered as the initial guess to the optimization solution for the new triplet being evaluated. This approach increases the likelihood of algorithm convergence for every feasible trim point and minimizes the risk of becoming trapped in local minima [37].

For commercial transport aircraft, bank angles of less than 30 degrees are preferred in the path planning, especially in the final approach and landing [7]. This allows for shallow turns that subject passengers to small $g$-forces (up to $1.2g$). It is also a common practice to impose such a constraint in studies involving a conventional civil airliner as the study model. For example, in [21], the manoeuvring flight envelopes of a wing damaged GTM were estimated with bank angles limited to $\pm 20°$, or a 35° bank constraint was imposed on the nominal and off-nominal flight envelope determination of a jet transport model in [27]. Therefore, a 30° bank constraint was considered in the derivation of the impaired commercial aircraft's MFEs utilized in the trajectory optimization approach of this research, as in (15). It should be noted that the stall constraint on the bank angle is observed in the MFEs used in this study. This is evident in Figures 1 and 2(a) by the gradual increase in the total airspeed value of the trim points at the lower boundary of the unimpaired MFE (i.e. stall speed) with the increase of the turn rate (bank angle). Furthermore, the structural loading constraint on the bank angle is met by adhering to the considered 30° bank limit.

Moreover, an additional angle of attack constraint was imposed to prevent the optimization algorithm from trimming the aircraft at the stall or post-stall flight conditions. As the NASA GTM was designed to investigate stall and post-stall regimes in the extended flight envelope regions, its aerodynamic dataset includes aerodynamic data for high values of angle of attack up to +85°. However, according to the results of the bifurcation analysis of the GTM presented in [4], from $\alpha = 10.5°$ the aircraft enters an undesirable regime with steep helical spirals which are considered as upset conditions and must be recovered from, by stall recovery procedures. Moreover, at the high angles of attack, there could be a shift away from the conventional flight dynamics modes, resulting in the controllers designed for normal flight to be less effective in the post-stall region. Such unconventional dynamics were studied in [69] for a T-tail configuration aircraft (i.e. NASA GTT). Hence, as shown in (16), the upper limit of $\alpha$ was constrained to $10.5°$ in the aforementioned trim point evaluation numerical procedure.

Another important factor taken into account is the maximum speed constraint due to structural loads and its relation with the lower limit of the flight path angle. Thrust-available and thrust-required curves have two intersections in the thrust-speed plane. At each specific altitude, the thrust-required curve varies with different flight path angles whilst the thrust-available curve is fixed [35]. Increasing the negative flight path angle (i.e. steeper descent) enlarges the weight component along the axis of the velocity vector, thus reducing the amount of the required thrust (i.e. lowering the thrust-required curve in the thrust-speed plane), and consequently increases the maximum speed. However, such higher speed should not exceed the aircraft's dive speed ($V_D$), to satisfy structural design limits. In the case of the GTM, information on dive speed is not provided. Therefore, as presented in (17), the lower limit of the flight path angle is constrained to $(-5°)$ and it is assumed that for the level flight ($\gamma = 0°$) and for the descending flight up to $(-5°)$, $V_{max}$ is sufficiently lower than the $V_D$. This assumption seems reasonable as $(-5°)$ flight path angle is in the range of typical descent angles of the commercial airliners.

Fig. 1 depicts 2D MFEs in the ($V - \dot{\psi}$) plane (i.e. slices of 3D MFEs at specific flight path angle) of an impaired GTM at 10000 ft and $(-5°)$ flight path angle. In this figure, MFE boundaries for rudder jamming cases of 0, 10, 20, and 25 degrees are indicated by black, red, cyan, and yellow dashed lines, respectively. Also, Figures 2(a) and 2(b) are two instances of the unimpaired and impaired 3D MFEs of the NASA GTM at constant altitude. Each of these flight envelopes consists of thousands of trim points. It is evident that various failure degrees (i.e. angles at which the rudder is jammed) result in completely different MFEs, emphasizing the need for a post-failure trajectory optimization approach that considers the remaining manoeuvrability of the impaired aircraft.

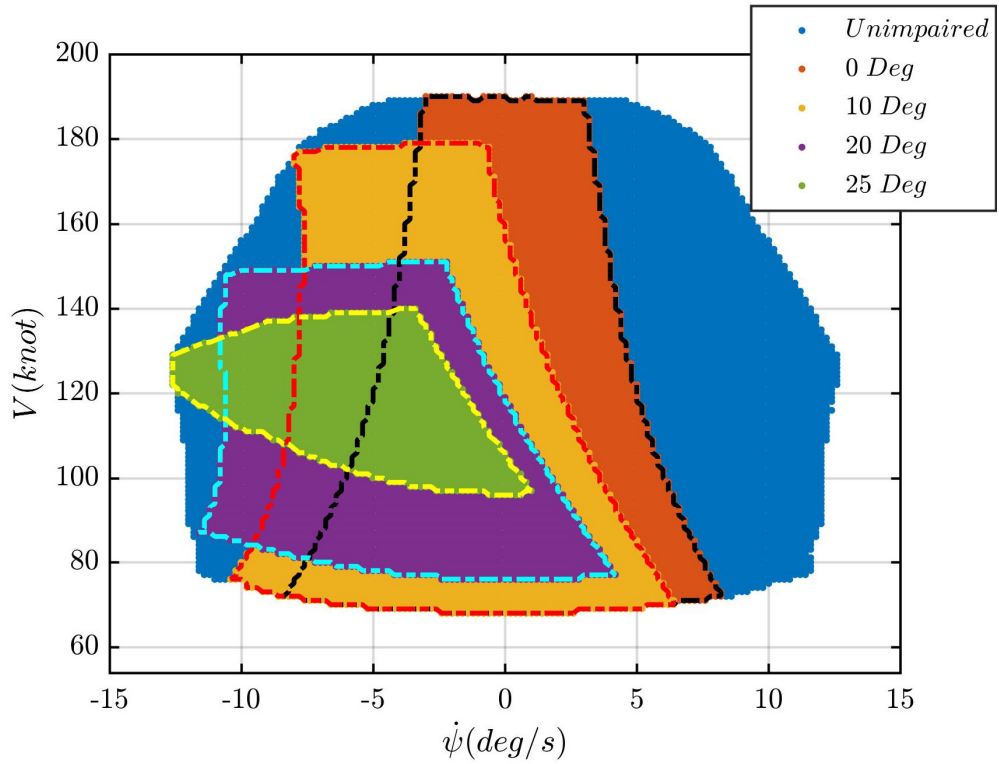

**Fig. 1 2D MFEs of five rudder jamming cases at 10000 ft and $\gamma = -5°$**

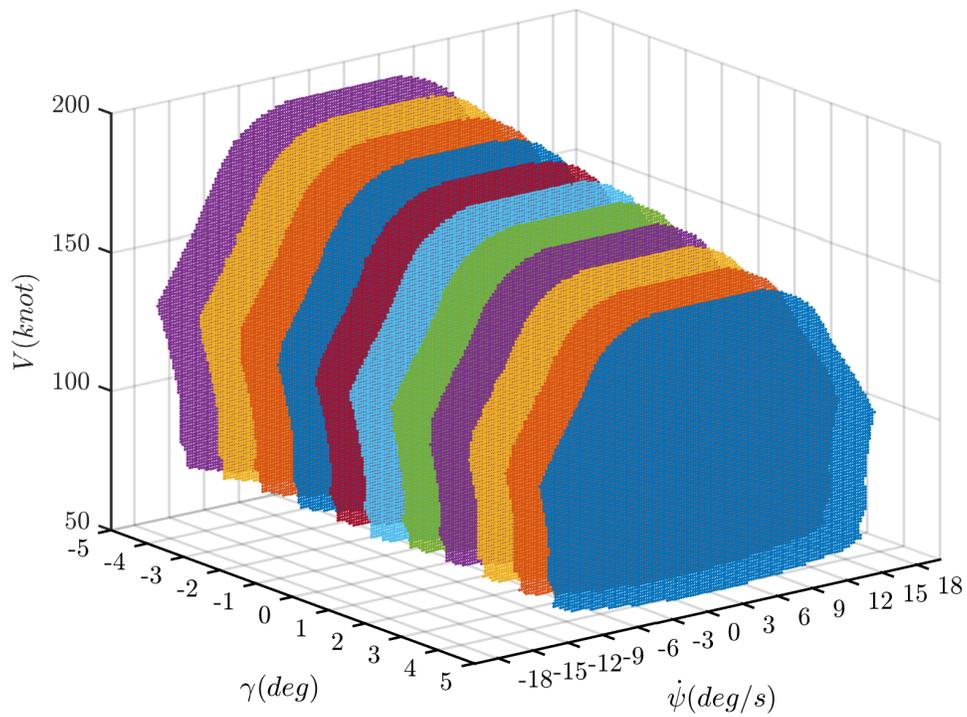

**a) Unimpaired case at sea level**

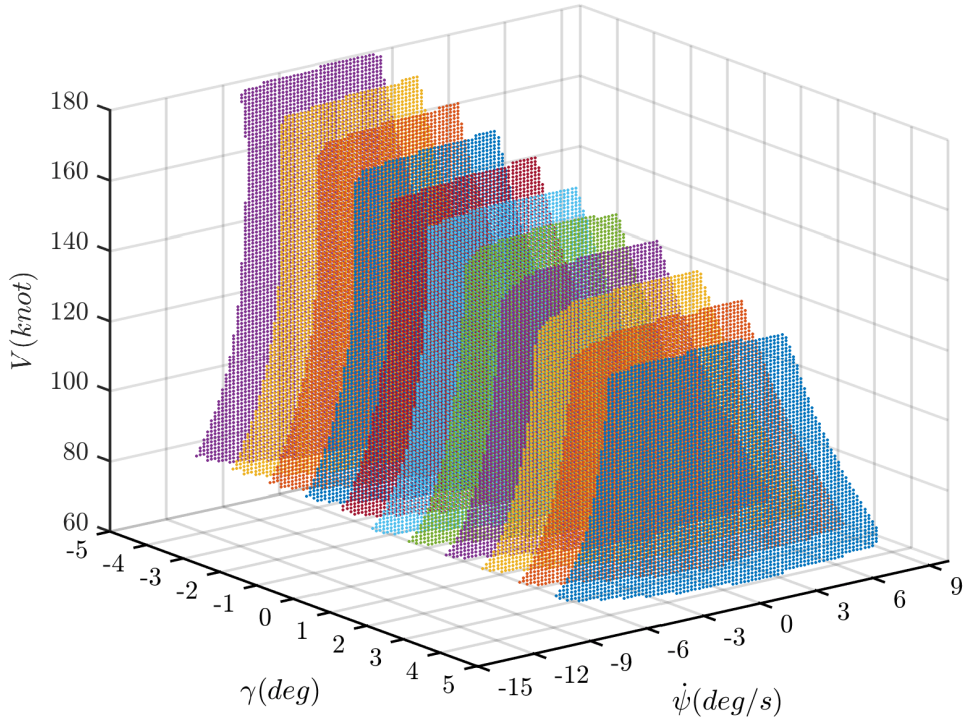

**b) Rudder jammed at 10° at 10000 ft**

**Fig. 2 Two instances of the 3D MFEs employed in this study**

## B. Path Planning via Optimal Control

The path planning or trajectory optimization problem can be effectively formulated as an optimal control problem, with historical roots in the calculus of variations dating back to the 17[th] century and initiated by Bernoulli's Brachistochrone problem and Newton's problem of Minimum Drag Revolution Surface in a Resisting Medium [70]. Detailed discussions on optimal control are covered in references [71-73]. The optimal control input is determined by minimizing the Hamiltonian, following Pontryagin's Maximum Principle (PMP).

Numerous path planning problems have been tackled using an optimal control approach over the past 60 years [74-84]. In optimal control problems, the vehicle dynamics are explicitly integrated into the formulation, thus ensuring the solution is not only dynamically feasible but also optimal, even if locally [85].

Formulating a trajectory optimization problem through the optimal control approach utilizing PMP results in a Two-point Boundary Value Problem (TBVP) [86]. Analytically solving the TBVP becomes challenging, particularly for nonlinear and constrained systems.

Another approach akin to PMP is Bellman's Principle of Optimality, serving as the foundation theorem for Dynamic Programming [87]. Its application to solve optimal control problems leads to the Hamilton-Jacobi-Bellman (HJB) equation, a partial differential equation (PDE). The analytical solution of the HJB equation also poses challenges.

Given the difficulties associated with obtaining analytical solutions for optimal control problems using HJB or PMP, numerical methods have been developed. These methods are broadly categorized into direct and indirect approaches. Indirect methods solve the optimal control problem by formulating the first-order optimality conditions, applying PMP, and employing Nonlinear Programming to obtain numerical solutions of the resulting Two-point Boundary Value Problem [88, 89]. However, these approaches are sensitive to initial guess, due to small convergence area. Additionally, they require the calculation of costate or adjoint variables, rendering them computationally expensive [86]. Consequently, these methods are not suitable for real-time implementation on aircraft [89]. Indirect methods include Indirect Shooting and Indirect Transcription [86].

On the contrary, direct methods have gained widespread popularity due to their suitability for real-time implementation. The fundamental concept behind these approaches is to transform the optimal control problem into a finite-dimensional nonlinear programming (NLP) problem by discretizing and parameterizing state and control variables. Subsequently, the NLP problem is solved using optimizers developed for this purpose. While direct methods may not guarantee an optimal solution compared to indirect methods, they often yield near-optimal solutions and have a larger convergence radius. Considering the inherent limitations of the overall system, due to disturbances, uncertainties, and atmospheric conditions, the primary focus for in-flight real-time trajectory optimization is on generating near-optimal paths within an acceptable timeframe. Therefore, direct methods are advantageous for this purpose [89].

Direct methods are categorized according to the subset of control and state variables they parameterize [90]. When only control and certain state variables are parameterized, explicit integration of the remaining state equations is required (e.g., Direct Shooting and Multiple Shooting) [91]. Conversely, when all state variables of the system are parameterized, explicit integration is not necessary (e.g., Pseudospectral [92, 93] and Inverse Dynamics [94-99] methods). The latter methods prove notably faster in comparison to integration methods [100].

Direct Shooting is suitable for problems with a small number of optimization variables because small changes introduced early in the trajectory can propagate into highly nonlinear changes at the end of the trajectory, resulting in nonlinear constraints. In the Multiple Shooting approach, time discretization is employed to tackle the issue of propagating changes, as any change at the beginning of each segment of the trajectory only affects that segment to its end, and the rest of the trajectory segments remain unaffected. However, ordinary differential equations (ODEs) defining dynamics of the system still need to be solved [86].

In order to avoid solving ODEs and increase optimization speed, in direct methods, differential equations are applied as dynamic constraints at the so-called collocation points (grid points) on the NLP variables (i.e. state and control variables). As ODEs are collocated (satisfied) as constraints at these points, they are referred to as the collocation points [86].

Direct methods depend on a specific grid for discretizing the original control problem. Enhancing the accuracy of the solution can be achieved by increasing the number of grid points. Despite the observed high solution accuracy with methods incorporating mesh refinement algorithms, implementing such approaches in real-time on aircraft is impractical due to the need for multiple iterations. In contrast, pseudospectral methods, which adhere to specific criteria for grid point selection, achieve higher accuracy compared to methods with a uniform grid and do not require consecutive iterations.

In pseudospectral [101-103] or orthogonal collocation [104, 105] methods, a finite basis of global interpolating polynomials is employed to approximate the state and control at a set of discretization points. In these methods, roots of an orthogonal polynomial (or the linear combinations of such polynomials and their derivatives) are selected as the collocation points; hence, pseudospectral methods are also called orthogonal collocation methods [106]. One advantage of pseudospectral methods is that, for smooth single-phase problems, they typically have faster convergence rates compared to other methods, exhibiting what is referred to as "spectral accuracy" [106, 107].

Pseudospectral methods stem from spectral methods traditionally used for solving fluid dynamics problems [102, 103]. Several well-known pseudospectral methods have been developed to solve optimal control problems in aerospace engineering, including the Chebyshev Pseudospectral Method (CPM) [108, 109] as utilized in [110], the Legendre Pseudospectral Method (LPM) [101] as used in [111] for the landing trajectory optimization, and the Gauss Pseudospectral Method (GPM) [112] as utilized in [113] for a quadrotor trajectory planning. CPM utilizes Chebyshev polynomials to approximate states and controls at Chebyshev-Gauss-Lobatto (CGL) points. A variant of CPM incorporating Clenshaw-Curtis quadrature is presented in [114]. LPM uses Lagrange polynomials to approximate variables and for orthogonal collocation at Legendre-Gauss-Lobatto (LGL) points [106].

In pseudospectral methods, the critical aspect significantly affecting the precision of these techniques is the selection of discretization points and collocation points. In many pseudospectral methods, these two types of points slightly differ. Discretization points (or nodes) are those where states are discretized, while collocation points are the points used to discretize controls and at which the differential equations and path constraints are satisfied as dynamic constraints. In some methods, in which collocation points exclude either one or both the initial and final points of the path, there is no distinction between discretization and collocation points except for the excluded points.

In direct methods, the necessary conditions in NLP, namely the Karush-Kuhn-Tucker (KKT) conditions, converge towards the optimal control necessary conditions, i.e. the Euler-Lagrange equations, as the number of variables increases, leading to improved accuracy. The Lagrange multipliers in NLP also converge towards the adjoint variables or costates in optimal control. If the solution obtained from NLP is an exact optimum (complete accuracy), the Lagrange multipliers in NLP will be equal to the costates in optimal control [86].

This principle establishes a criterion for assessing the accuracy of pseudospectral methods. Among various pseudospectral methods, in the GPM method, the solution obtained from the direct approach (NLP) also satisfies the first-order optimality conditions, akin to what an indirect method would yield [112]. This signifies the high accuracy of the GPM method. Furthermore, research conducted on both the LPM and RPM (Radau Pseudospectral Method) methods demonstrate rapid convergence rates of the NLP solution to the solution of the continuous-time optimality conditions in these methods [115-117]. The RPM method utilizes Legendre-Gauss-Radau points [115,118,119]. Research findings indicate that both the GPM and RPM methods have high accuracy, along with high computational speed and ease of implementation, whereas the LPM method is not as accurate as the other two methods [106].

In contrast to implicit integration methods such as Trapezoidal or Hermit-Simpson, which enforce defect constraints at arbitrary (irregular) discretization points and their midpoints, pseudospectral methods directly enforce constraints at collocation points, which are roots of orthogonal polynomials. Consequently, the derivatives of state variables can be easily calculated through the differentiation matrix derived in these methods [120]. Thus, pseudospectral methods facilitate fast and accurate numerical interpolation, differentiation, and integration [121].

In this research, the optimal control problem of trajectory optimization is efficiently transformed into NLP, and then Radau pseudospectral method is employed to solve it.

## C. Aircraft Dynamic Model

The emergency approach path planning for an impaired aircraft must yield an optimal or near-optimal solution, ensure terrain avoidance, and be implemented in real-time. Studies have shown that achieving real-time trajectory optimization with a 6-DoF model faces challenges due to the substantial number of state and control variables. The necessity of applying differential defect constraints to these variables at each collocation point significantly increases the number of NLP decision variables, making real-time results unattainable.

For instance, a recent study in [122] proposed a trajectory optimization approach for an eVTOL (electric vertical take-off and landing) aircraft that incorporated 6-DoF flight dynamics. The approach used trapezoidal collocation to transcribe the optimal control problem into an NLP, which as mentioned earlier, is neither as fast nor as accurate as pseudospectral methods. Additionally, the approach did not take terrain into account and, given the flight times of the generated trajectories, the computational times for the optimization were beyond real-time applications. Furthermore, the trajectory optimization only considered nominal flight envelope constraints, which do not account for in-flight failure scenarios.

Additionally, employing a 6-DoF model, coupled with an increase in the number of nodes, results in an expansion of the constraints matrix dimensions in the optimization problem. This expansion may surpass the capacities of certain available solvers, rendering the problem-solving process infeasible.

On the other hand, geometrical (kinematic) path planning, which disregards the aircraft dynamics, may not necessarily be feasible and is most likely not optimal, despite being implementable in real-time.

Even, online 3D trajectory generation methods proposed recently such as [123] do not incorporate high-fidelity nonlinear 6-DoF models as they are aimed at generating deterministic trajectories online with a low computational cost for aircraft certification purposes, rather than considering the optimality of the generated path. As a result, they are not suitable for impaired aircraft trajectory optimization.

Fortunately, by considering the point-mass aircraft model (3-DoF), it is possible to consider the aircraft's coordinate variables $(x, y, z)$ as the system's flat outputs, as the other state and control variables can be derived from these three variables [47, 84]. In this case, the system can be considered differentially flat, which significantly helps in reducing the number of decision variables in the optimization process of NLP and prevents further complications of boundary conditions and constraints resulting from choosing output variables instead of state and control variables [89].

The aforementioned approach (i.e. combination of the inverse dynamics technique with the 3-DoF model) has been employed in [48, 50, 89, 124], representing one of the highly cited real-time trajectory optimization methods. In these studies, each iteration of trajectory optimization involves the calculation of the three-dimensional trajectory, followed by the computation of the aircraft's body angular velocities through quaternions differentiation. Then, the nondimensional moment coefficients are determined, and the values of the control variables of the aircraft are evaluated. Subsequently, the values of the control parameters (control surfaces, throttle, and pseudo-controls ($\alpha$ and $\phi$)) are checked within their associated constraints, and the cost function is computed. This iterative process continues until the optimal solution is reached.

In [48], this method was utilized to generate an optimal terrain following and terrain avoidance (TF/TA) three-dimensional path for a nonlinear 6-DoF model. It should be noted that the time required to generate the single-piece optimal trajectory cannot be considered real-time, and when considering the path traverse time, it can't be neglected. However, suboptimal trajectories can be generated rapidly, in a fraction of a second.

Nevertheless, the aforementioned approach is not applicable to most impaired aircraft due to the assumption of zero sideslip angle ($\beta = 0$). In fact, the success of the aforementioned method in generating real-time suboptimal three-dimensional trajectories for a nonlinear 6-DoF model relies on the assumption of ($\beta = 0$) throughout the path. This assumption implies that the aircraft performs non-rectilinear manoeuvres along the path in a coordinated manner. However, in most cases of aircraft structural damage or control surfaces failure (as extensively discussed in our previous work [34]), due to the continuous incurred moment caused by the impaired aircraft's geometric asymmetry or by the damaged control surface, the aircraft is constantly sideslipped. Thus, the assumption of ($\beta = 0$) is not valid for such impaired aircraft cases.

The above point can be made clearer by considering the following 3-DoF equations of motion of the aircraft:

$$\dot{x} = V \cos\gamma \cos\chi \tag{18}$$

$$\dot{y} = V \cos\gamma \sin\chi \tag{19}$$

$$\dot{z} = -V \sin\gamma \tag{20}$$

$$\dot{V} = \frac{(T \cos\alpha - D)}{m} - g \sin\gamma \tag{21}$$

$$\dot{\chi} = [(-T \cos\alpha \sin\beta)\cos\phi + (T\sin\alpha + L)\sin\phi]/mV\cos\gamma \tag{22}$$

$$\dot{\gamma} = [(-T \cos\alpha \sin\beta)\sin\phi - (T\sin\alpha + L)\cos\phi]/mV - g \cos\gamma/V \tag{23}$$

In the above equations, $\chi$ represents the heading angle of the aircraft in velocity axes, and its magnitude differs from $\psi$ (yaw angle), whose derivative equates to the aircraft turn rate. In fact, based on the relationship between body, wind, velocity, and inertial axes systems, while $\psi$ is one of the three angles used to transform inertial axes to body axes, $\chi$ is used along with the flight path angle ($\gamma$) to transform inertial axes to velocity axes (refer to Fig. 2.2-7 in [125], wherein $\chi$ is denoted by $\xi$, and corresponding discussions therein). The relationship between $\chi$, $\psi$, and $\beta$ can be expressed as follows [125]:

$$\chi = \psi + \beta \tag{24}$$

In this study, one of the four parameters characterizing trim points is $\dot{\psi}$, not $\dot{\chi}$. According to equation (24), $\dot{\chi}$ depends on $\dot{\psi}$ and $\dot{\beta}$. Since there isn't any direct relationship upon which $\beta$ can be formulated in velocity axes, and the equations for $\dot{\psi}$ and $\dot{\beta}$ in body and wind axes depend on the values of the angular velocities $(q, r)$ and $(p, r)$, respectively, as per equations (25) and (26), $\beta$ or $\dot{\beta}$ cannot be replaced in the aircraft 3-DoF equations of motion.

$$\dot{\psi} = (q \sin \phi + r \cos \phi) \sec \theta \tag{25}$$

$$\dot{\beta} = \frac{g}{V} (\cos \beta \sin \phi \cos \theta + \sin \beta \cos \alpha \sin \theta - \sin \alpha \cos \phi \cos \theta \sin \beta)$$

$$+ p \sin \alpha - r \cos \alpha + \bar{q}S/_{mV} (C_Y \cos \beta + C_D \sin \beta) + T/_{mV} \cos(\alpha + \sigma_T) \sin \beta \tag{26}$$

This is why various studies employing 3-DoF equations for flight simulation or trajectory optimization have not considered non-zero values of $\beta$, assuming $\chi = \psi$. Therefore, the 3-DoF equations of motion (18) – (23) cannot be applied to the case of an impaired aircraft in this research. The utilization of the 12 6-DoF equations of motion in wind axes also evidently renders real-time trajectory optimization infeasible.

To tackle this challenge, this research adopts an approach based on a pre-computed database of manoeuvring flight envelopes. Unlike the traditional approach to aircraft path planning, which involves calculating the required values of control inputs to achieve the trajectory; during path planning using available methods, this approach capitalizes on the fact that the main and time-consuming part of path planning computations were previously done during the generation of the manoeuvring flight envelopes database.

In other words, for each impaired aircraft, it has already been calculated what manoeuvres it can execute at various flight altitudes while adhering to all dynamic constraints. Therefore, the problem can be reduced to finding a terrain-avoiding path $(x, y, z)$ composed of triplets $(V^*, \gamma^*, \dot{\psi}^*)$, where each trim point in the manoeuvring flight envelope is characterized by such a triplet. In this method, instead of searching for permissible control parameter values and aerodynamic angles during the optimization iterations, only the values of $(V^*, \gamma^*, \dot{\psi}^*)$ are pursued which are inside the impaired aircraft's manoeuvring flight envelope and their corresponding three-dimensional path would lead the aircraft to the landing runway without any collision with terrain.

Following this approach, the challenge posed by the 3-DoF equations of motion with an existing sideslip angle can be addressed as follows: each manoeuvring flight envelope of an impaired aircraft is essentially a subset of the nominal manoeuvring flight envelope of an unimpaired aircraft. In other words, every steady manoeuvre achievable by the impaired aircraft (i.e., a feasible, stable, and controllable trim point [34, 36]) is also executable by the unimpaired aircraft. The key distinction lies in the fact that the impaired aircraft (e.g. when consistently sideslipped due to the incurred damage) executes the manoeuvre with different values of control parameters and aerodynamic angles compared to those of the unimpaired aircraft. Despite these differences, the path generated by that manoeuvre remains the same in 3D space for both the impaired and unimpaired aircraft.

For instance, in Fig. 3, all manoeuvres within the manoeuvring flight envelope of an impaired GTM with jammed rudder at 10 degrees (the red flight envelope) are also encompassed by the boundaries of the unimpaired aircraft envelope at sea level (the blue flight envelope). However, the values of the control surfaces, throttle, and angle of attack needed for the same manoeuvres (i.e., identical $(V^*, \gamma^*, \dot{\psi}^*)$) differ between the two flight envelopes. Furthermore, while each manoeuvre within the unimpaired flight envelope can be executed without a sideslip angle, the same manoeuvre, in the case of a jammed rudder, necessitates a sideslip angle. This is because the positive rudder jamming angle induces a negative yawing moment, steering the aircraft's nose to the left. The positive sideslip angle counteracts the effect of the jammed rudder by generating a non-zero side force, resulting in a positive yawing moment.

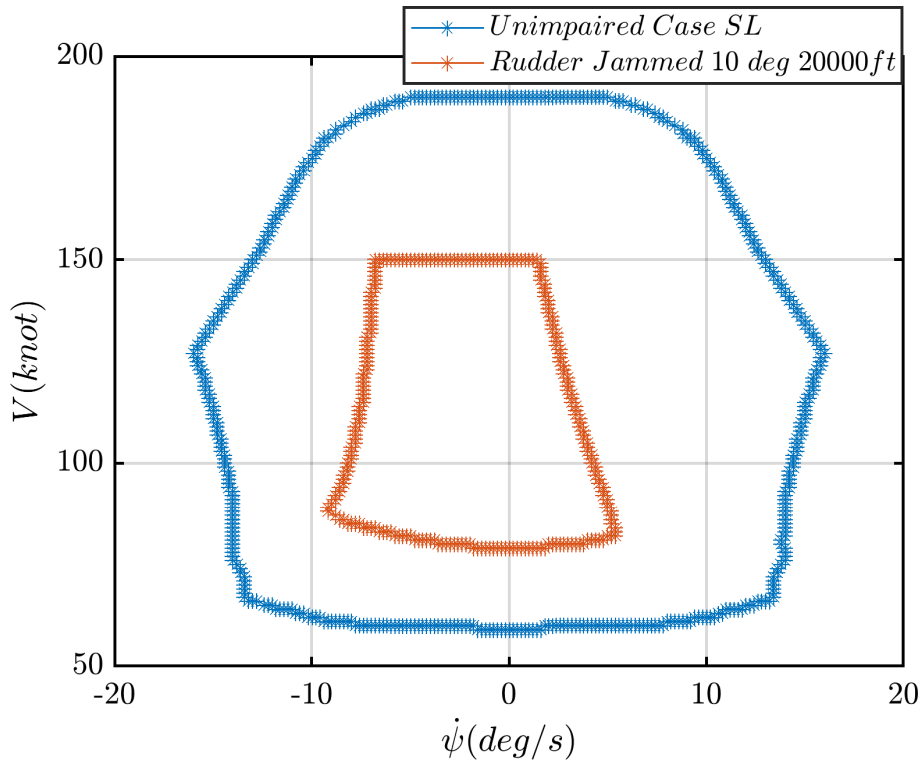

**Fig. 3 Unimpaired and 10° jammed rudder MFEs of NASA GTM at different altitudes and $\gamma = 0°$**

As the generated path remains the same with the identical sequence of the same trim points, whether these trim points are chosen from the manoeuvring flight envelope of the impaired aircraft or the nominal manoeuvring flight envelope of the unimpaired aircraft, and only the values of the control parameters and aerodynamic angles required for path planning differ between the two scenarios, the aircraft's equations of motion can be reformulated as follows:

$$\dot{x} = V \cos\gamma \, \cos\psi \qquad (27)$$

$$\dot{y} = V \cos\gamma \, \sin\psi \qquad (28)$$

$$\dot{z} = -V \sin\gamma \qquad (29)$$

Essentially, these equations mirror the first three equations in the 3-DOF equations of motion (i.e., (18)-(20)), with the substitution of $\chi$ for $\psi$. This substitution arises from the fact that manoeuvres within the unimpaired flight envelope can be executed without the need for sideslip. Simply put, the path planning algorithm (NLP solver) doesn't need to be informed of the damage, and only has to select trim points from the unimpaired flight envelope (depicted by the blue boundary in Fig. 3) ensuring they also fall within the red boundary (representing the impaired aircraft flight envelope in Fig. 3). This guarantees that the impaired aircraft can execute these manoeuvres, albeit with sideslip and different control parameter values. In essence, the manoeuvring flight envelope of the impaired aircraft signifies that all essential calculations for manoeuvres within its boundaries have been previously conducted. Consequently, there's no need for the path planning algorithm to re-compute whether the control parameters and aerodynamic angles required for manoeuvres within the envelope satisfy the dynamic constraints. Therefore, immediately after the occurrence of failure:

- Boundary of the impaired aircraft's manoeuvring flight envelope is estimated in real-time using the neural network-based approach presented in [36].

- It is assumed that all attainable manoeuvres of the impaired aircraft are inside the unimpaired aircraft's flight envelope, and the three equations (27)-(29) are employed as the dynamic model of the path planning problem considering $x, y, z$ and $\psi$ as state variables and $V$, $\gamma$, $\dot{\psi}$ as control inputs.

- The values of trim points that are on the boundary of the impaired aircraft's manoeuvring flight envelope, along with the safe altitude requirements for terrain clearance, are considered as the path constraints in the problem.

- Optimizer selects trim points that lead the impaired aircraft safely to the runway, based on the considered optimization criteria. For the failure scenarios including a damaged control surface, it is suggested that the trim point selection is aimed at the minimum control effort. This is to avoid excessive control effort possibly causing a secondary actuator failure that could lead to LOC [65], and also to keep the aircraft in vicinity of trim conditions; ensuring minimum discomfort for passengers.

- Subsequently, the control parameters and aerodynamic angles are swiftly computed (in less than a second) at the chosen trim points, considering the impact of the incurred damage on the aircraft dynamics. In other words, the impaired aircraft is promptly trimmed at the selected points to obtain the specific values of the required control parameters and aerodynamic angles corresponding to those trim points.

Therefore, the optimal control problem of post-failure emergency landing trajectory optimization can be formulated as the following. The cost function

$$J = \int_{t_0}^{t_f} [\mathbb{x}(t), \mathbb{u}(t), t] \, dt \tag{30}$$

should be minimized subject to dynamic constraints

$$\dot{\mathbb{x}}(t) = f[\mathbb{x}(t), \mathbb{u}(t), t], \quad t \in [t_0, t_f] \tag{31}$$

where

$$\mathbb{x} = [x, y, z, \psi]^T \tag{32}$$

$$\mathbb{u} = [V, \gamma, \dot{\psi}]^T \tag{33}$$

and (31) is the compact form of the equations (27)-(29). The optimization is also subject to path constraints

$$\text{MFE}_{LL} \leq \mathbb{u}(t) \leq \text{MFE}_{UL} \tag{34}$$

$$[h = -z] \geq h_{Terr} + h_{Cl} \tag{35}$$

where $\text{MFE}_{UL}$ and $\text{MFE}_{LL}$ are 3-dimensional vectors representing the upper and lower permissible limits of $V$, $\gamma$, $\dot{\psi}$ based on the boundaries of the 3D manoeuvring flight envelope at each specific altitude. Such limits are derived by nonlinear interrelation between aircraft dynamic parameters and can be depicted via 3D MFEs or 2D ($V - \dot{\psi}$) MFEs at constant flight path angles. For instance, at specific trim points, attainable turn rate is limited by stall speed which is dependent on aircraft bank angle, sideslip angle and angle of attack, whereas at other trim points the limiting factor could be aileron saturation. Comprehensive and detailed discussions on limiting parameters of MFE boundaries are presented in [34].

(35) defines the aforementioned safe altitude requirements over terrain, wherein, flight altitude $h$ must be greater than the sum of the terrain height $h_{Terr}$ and the safe clearance altitude $h_{Cl}$.

Once the trajectory trim points (i.e. triplets $(V, \gamma, \dot{\psi})$) are selected by the optimizer, their corresponding control inputs and aerodynamic angles are computed via trimming the impaired aircraft at the selected trim points through the constrained nonlinear optimization process presented in (10)-(17).

## D. Radau Pseudospectral Method

The optimal control problem presented in (30)-(35) can be solved using the Radau pseudospectral method (RPM). The RPM is an orthogonal collocation method wherein the collocation nodes are the Legendre-Gauss-Radau (LGR) points $(\tau_1, \tau_2, \ldots, \tau_N)$ defined on the interval $-1$ to $+1$ such that $\tau \in [-1, +1)$, $\tau_1 = -1$ and $\tau_N < +1$. An additional non-collocated point $\tau_{N+1} = +1$ is used to describe the approximation to the state variable [126]. Therefore in order to accomplish the full discretization of the time interval, the $N + 1$ discretization points are found utilizing the $N$ LGR points plus the end point $\tau_{N+1} \equiv +1$. It is noteworthy that unlike other pseudospectral methods such as the Legendre pseudospectral method (LPM), in the RPM different number of points are used for collocation and discretization. Specifically, the collocation points are the $N$ LGR points, whilst the discretization points are the LGR points plus the end point $\tau_{N+1}$ [106, 127]. Assuming $\mathcal{L}_i(\tau), (i = 1, \ldots, N + 1)$ to be a basis of $N + 1$ Lagrange polynomials with support points at $(\tau_1, \tau_2, \ldots, \tau_{N+1})$ and given by

$$\mathcal{L}_i(\tau) = \prod_{\substack{j=1 \\ j \neq i}}^{N+1} \frac{\tau - \tau_j}{\tau_i - \tau_j}, \quad (i = 1, \ldots, N + 1) \tag{36}$$

then the state is approximated by a series of the form

$$\mathbb{x} \approx \mathcal{X}(\tau) = \sum_{i=1}^{N+1} \mathcal{L}_i(\tau) \, \mathcal{X}(\tau_i) \tag{37}$$

Differentiating and evaluating the series at the collocation point $\tau_k$ gives

$$\dot{\mathbb{x}} \approx \dot{\mathcal{X}}(\tau_k) = \sum_{i=1}^{N+1} \dot{\mathcal{L}}_i(\tau_k) \, \mathcal{X}_i(\tau_k) = \sum_{i=1}^{N+1} \mathcal{D}_{ki} \, \mathcal{X}_i(\tau_k) \tag{38}$$

where $\mathcal{D}$ is a $N$ by $N + 1$ differentiation matrix comprising one row for each collocation point and the derivatives of the Lagrange polynomials evaluated at each of the collocation points in the columns [126], and defined as

$$\mathcal{D}_{ki} = \dot{\mathcal{L}}_i(\tau_k) = \begin{cases} \dfrac{\dot{g}(\tau_k)}{(\tau_k - \tau_i)\dot{g}(\tau_i)}, & \text{if } k \neq i \\ \dfrac{\ddot{g}(\tau_i)}{2\dot{g}(\tau_i)}, & \text{if } k = i \end{cases} \tag{39}$$

where $g(\tau_i) = (1 + \tau_i)[\mathcal{P}_{N+1}(\tau_i) - \mathcal{P}_N(\tau_i)]$, $\mathcal{P}_N$ is the $N$th degree Legendre polynomial, and $(\tau_i)$ are the $N$ LGR points plus the end point [106]. The derivatives can be expressed as

$$\dot{\mathcal{X}} \approx \frac{1}{\varsigma} \mathcal{D} \hat{\mathcal{X}} = \frac{d\mathcal{X}}{d\tau} = \mathfrak{f}\big(\mathcal{X}(\tau), \mathcal{U}(\tau)\big) \tag{40}$$

where $\hat{\mathcal{X}}$ is the discretized state vector across all nodes, $\mathfrak{f}$ is a vector of nonlinear functions representing dynamic constraints in the computational time domain $\tau$, and $\varsigma$ is the transformation metric between computational time domain $\tau$ and physical time domain $t$, defined as

$$t = \frac{(t_f - t_0)\tau}{2} + \frac{(t_f + t_0)}{2} \tag{41}$$

$$\varsigma \triangleq \frac{dt}{d\tau} = \frac{(t_f - t_0)}{2} \tag{42}$$

Therefore, the collocation equations can be written as

$$\sum_{i=1}^{N+1} \mathcal{D}_{ki} \, \mathcal{X}(\tau_i) - \frac{t_f - t_0}{2} \mathfrak{f}\big(\mathcal{X}(\tau_k), \mathcal{U}(\tau_k)\big) = 0, \quad (k = 1, \dots, N) \tag{43}$$

Assuming $\mathcal{W}_i, 1 \le i \le N$ are the quadrature weights associated with the LGR points, these weights have the property that for any function $\mathcal{S}$

$$\int_{-1}^{1} \mathcal{S}(\tau)d(\tau) = \sum_{i=1}^{N} \mathcal{W}_i \mathcal{S}(\tau_i) \tag{44}$$

Hence, the cost function can be discretized as follows:

$$\int_{t_0}^{t_f} \mathbb{F}(\tau)d(t) = \varsigma \int_{-1}^{1} \mathbb{F}(t(\tau))d(\tau) \approx \varsigma \sum_{i=1}^{N} \mathcal{W}_i \mathbb{F}(t_i) \tag{45}$$

Eventually, as the collocation equations involve the control solely at the LGR points, the control is approximated using $N$ Lagrange polynomials $\bar{\mathcal{L}}_k(\tau), (k = 1, \dots, N)$ as

$$\mathbb{u} \approx \mathcal{U}(\tau) = \sum_{k=1}^{N} \bar{\mathcal{L}}_k(\tau) \, \mathcal{U}(\tau_k) \tag{46}$$

The explanation above provides a concise overview of the use of the Radau pseudospectral method in solving an optimal control problem. For further details, refer to [106, 126, and 127].

The aforementioned approach transforms the continuous optimal control problem into a discrete parameter optimization problem. The resulting nonlinear programming problem is then solved using an appropriate solver. As will be discussed in the following section, for the case study of this research, the RPM was applied to the impaired aircraft's trajectory optimization problem using GPOPS (General Pseudospectral Optimization Software) within MATLAB, and the NLP was solved using the MATLAB implementation of SNOPT (Sparse Nonlinear OPTimizer) solver.

# III.     Numerical Simulation and Discussion of Results

A key element in every scientific research is a valid model. The case study model used in this research is based on the NASA Generic Transport Model (GTM), which resembles a commercial airliner. The GTM – with tail number T2, is a 5.5% twin – turbine powered, dynamically scaled aircraft which is designed with the aim of flying into drastic upset conditions and being safely recovered. Extensive wind tunnel tests were performed on the GTM to create an extended-envelope aerodynamic data set. Test data were obtained at angles of attack as low as $-5°$ and up to $+85°$ and sideslip angles ranging from $-45°$ to $+45°$ [128]. The GTM-T2 properties are shown in Table 1. It should be noted that the influence of damage cases in the GTM mathematical model involves both aerodynamic and inertial effects.

**Table 1 GTM-T2 Properties**

| Property | Quantity |
|---|---|
| Takeoff weight, $W_0$ | 257 N (26.2 kg) |
| Wing area, S | 0.5483 m$^2$ |
| Wing span, d | 2.09 m |
| Length, l | 2.59 m |
| Mean aerodynamic chord, $\bar{c}$ | 0.2790 m |

## A.  Optimization Criterion

The cost function utilized in the case study of this research, which adopts a minimum control effort approach, can be expressed in a Lagrange formulation as below. This approach optimizes the trajectory by reducing the use of control inputs, leading to smoother, more efficient, and safer manoeuvres:

$$J = \int_{t=0}^{t=tf} \mathbb{u}^T(t) R(t) \mathbb{u}(t) dt \tag{47}$$

where $R$ is a weighing matrix. It should be noted that as mentioned previously, the control vector in (47) is as presented in (33), in accordance with the proposed approach of this study. Selecting $V$, $\gamma$ and $\dot{\psi}$ as control inputs has dual advantages. It enables the direct selection of the trim parameters, expediting the NLP problem-solving process significantly. Simultaneously, minimizing them as pseudocontrols leads to the minimum effort of control variables $\delta_{th}$, $\delta_e$, $\delta_a$, $\delta_r$. This is because $\dot{\psi}$ is a pseudocontrol that reflects the effect of $\delta_a$ and $\delta_r$, and $V$ effectively represents $\delta_{th}$, while $\gamma$ is influenced by both $\delta_{th}$ and $\delta_e$. This approach has also been employed in other studies that have used a point mass model, with the difference being the presence of $\alpha$ and $\phi$ in the equations for which these two parameters are considered as longitudinal and lateral pseudocontrol inputs alongside the thrust.

## B. Numerical Solver

Various solvers are available to solve the trajectory optimization NLP problem, for which the transcription code must be written, and the problem must be modelled. These solvers are then utilized to solve the formulated problem. Three commonly employed solvers include SNOPT, FMINCON, and IPOPT, all of which offer compiled codes for use in MATLAB. While SNOPT and IPOPT are designed for solving sparse problems, they can handle dense problems of small to medium sizes. SNOPT is based on the SQP method, while IPOPT is based on the Interior-Point method. One notable advantage of SNOPT is its warm-start capability, attributed to the SQP method. Consequently, SQP – based methods, particularly SNOPT, are widely used in solving path planning problems. FMINCON, which is a built – in MATLAB function, also supports SQP but performs less effectively in finding the true optimal solution compared to SNOPT. This is because in SNOPT a single function call is used for the objective function and all constraints, while FMINCON requires that the objective function and constraint function are given in separate functions, leading to the constraints and objective function re-evaluating the dynamics at every grid point, thus reducing computational speed [121].

In this research, GPOPS (General Pseudospectral Optimal Control Software), a general – purpose MATLAB software for solving optimal control problems, is used to solve the trajectory optimization problem. It models the optimal control problem as an NLP problem based on the Radau Pseudospectral Method (RPM) and then utilizes SNOPT to solve it. Considering the high convergence speed and accuracy of the RPM, and also GPOPS' well – structured interface and its internal connection with SNOPT, this software is one of the best choices for solving optimal control problems including trajectory optimization [129].

## C. Terrain Modelling

In the conducted case study of this research, it is assumed that at the flight altitude of 6000 metres, GTM incurs a rudder failure of jamming at 10 degrees. The aircraft needs to reach the landing runway at sea level (zero altitude), positioned 16000 metres longitudinally and 19000 metres laterally away from the point of the occurrence of the damage. However, in between there is a set of terrain obstacles with a maximum height of 7000 metres that must be avoided. Unlike similar studies that employed simpler terrain or chose less challenging terrain altitude and location with respect to the aircraft flight path, this study deliberately selected a more intricate terrain set and runway placement to rigorously evaluate the proposed method's efficacy.

In this research, the MSG (Max-Set of Gaussians) landscape generator proposed in [130] is used to model the intended terrain as a combination of $n$-dimensional Gaussian functions. It is important for the modelled terrain to be smooth and differentiable. Terrain model generated using Digital Elevation Map (DEM) data, although accurately representing the real – world terrain, is not differentiable. Therefore, sometimes trigonometric functions are used in related research as an alternative since they are smooth and differentiable, but terrain models generated with such functions are highly unrealistic. In this study, Gaussian functions of the following form that are differentiable are used as the fundamental component of the landscape generator:

$$g(\mathfrak{X}) = [\frac{1}{(2\pi)^{\frac{n}{2}}|\Sigma|^{\frac{1}{2}}}\exp\left(-\frac{1}{2}(\mathfrak{X} - \mu)\Sigma^{-1}(\mathfrak{X} - \mu)^T\right)]^{1/n} \tag{48}$$

wherein $\mu$ and $\Sigma$ represent the $n$-dimensional vector of means and the $(n \times n)$ covariance matrix, respectively, and $\mathfrak{X}$ is the vector of sampling points in the search space. Such a Gaussian function is the basic building block of the landscape generator. The landscape contains several Gaussian functions, each constituting a "bump" or "hill" in the landscape [131]. Moreover, in order to avoid collision with the generated terrain, a path constraint is applied to the problem based on existing regulations, requiring the aircraft to maintain a minimum of 2000 feet separation with terrain (minimum obstacle clearance altitude denoted by $h_{cl}$ in (35)) in mountainous areas (i.e. areas having a terrain elevation differential exceeding 3,000 feet within 10 nautical miles) [132].

**D. Case Study Results**

In this case study, first an optimal one-piece path with high accuracy (considering a high number of nodes) is generated but due to the path's length and the complexity of the terrain, the generation time exceeds real-time constraints. Additionally, a near-optimal piecewise path consisting of several optimal trajectories is generated in real-time. These two paths and their specifications, especially in terms of optimality (considering the cost function values of each path), are compared. The results indicate that the time required to generate each of the 30 optimal trajectory segments of the near-optimal piecewise path is approximately a fraction of a second to slightly over a second, enabling real-time implementation during path traversal.

For the one-piece optimal trajectory, given the expansive search space and the applied path constraints, particularly the mountainous terrain, it is unlikely for the algorithm to yield a solution when attempting to solve the NLP problem from the start with a large number of nodes. This is because the optimization process progresses very slowly, and there's a risk of reaching the maximum allowable iterations of SNOPT before the problem is entirely solved. Therefore, in this study, the problem is initially solved with 20 nodes, and then this initial solution is used as an initial guess to solve the problem with 100 nodes. This process is repeated iteratively: solving with 100 nodes and providing the solution for the problem with 200 nodes, solving with 200 nodes and providing the solution for problem with 300 nodes, and eventually, solving with 300 nodes and providing the corresponding solution for the problem with 400 nodes. At each step that the number of nodes is increased, the algorithm refines the path, especially the path segments that were clear of obstacles at a lower number of nodes (e.g. a mountain peak located between two nodes) but have some nodes going through terrain at a higher number of nodes, which are adjusted according to the cost function and other problem constraints.

It is noteworthy that the manoeuvring flight envelope of the impaired aircraft can be evaluated in real-time at any intended altitude using the real-time approach proposed in our previous research [36]. Additionally, as discussed and analysed in detail in [34] regarding the variations in manoeuvring flight envelope with flight condition, manoeuvring flight envelopes expand as altitude decreases, thereby increasing the number of attainable manoeuvres by the impaired aircraft. This expansion facilitates the algorithm's problem-solving process. However, in this case study, to assess the robustness of the proposed algorithm, the boundaries of the more restricted impaired aircraft's flight envelope at 6000 metres have been upheld as path constraints until the trajectory reaches sea level.

Figs. 4-6 show different projections of the generated 3D trajectories and the terrain. In these figures, the red path depicts the one-piece optimal trajectory, while the magenta path represents the near-optimal piecewise trajectory comprising 30 optimal segments. Moreover, Fig. 7 is a contour plot presenting the height of the 3D terrain across the 2D landscape using the vertical colour bar.

Corresponding time histories plots of the one-piece non-real-time trajectory and the piecewise real-time trajectory are presented in Figures 8 to 13, and 14 to 19, respectively. Specifically, the variation of linear velocity components over time is depicted in Figures 8 and 14, while Figures 9 and 15 show the variation of angular velocity components along the trajectory. Figures 10 and 16 present the variation of total airspeed, angle of attack, and sideslip angle, while Figures 11 and 17 illustrate the variation of Euler angles rates. Figures 12 and 18 display the variation of Euler angles and flight path angle over time, and Figures 13 and 19 present the variation of control inputs.

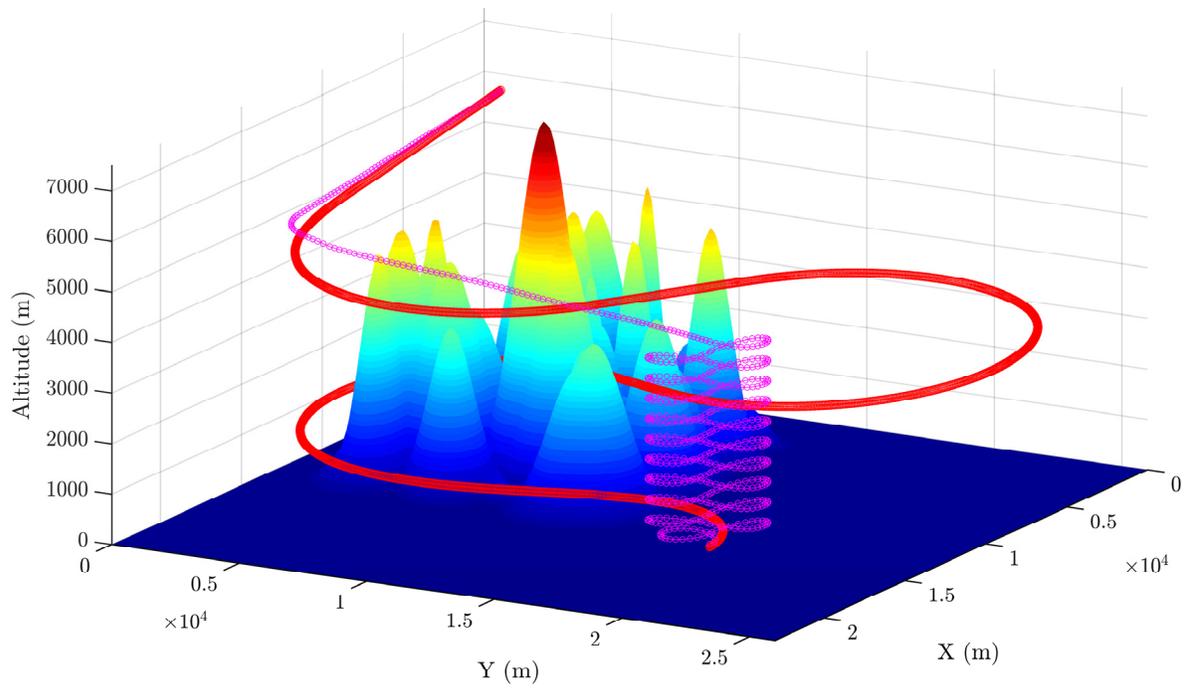

**Fig. 4 3D trajectories generated for impaired GTM with jammed rudder at 10°**

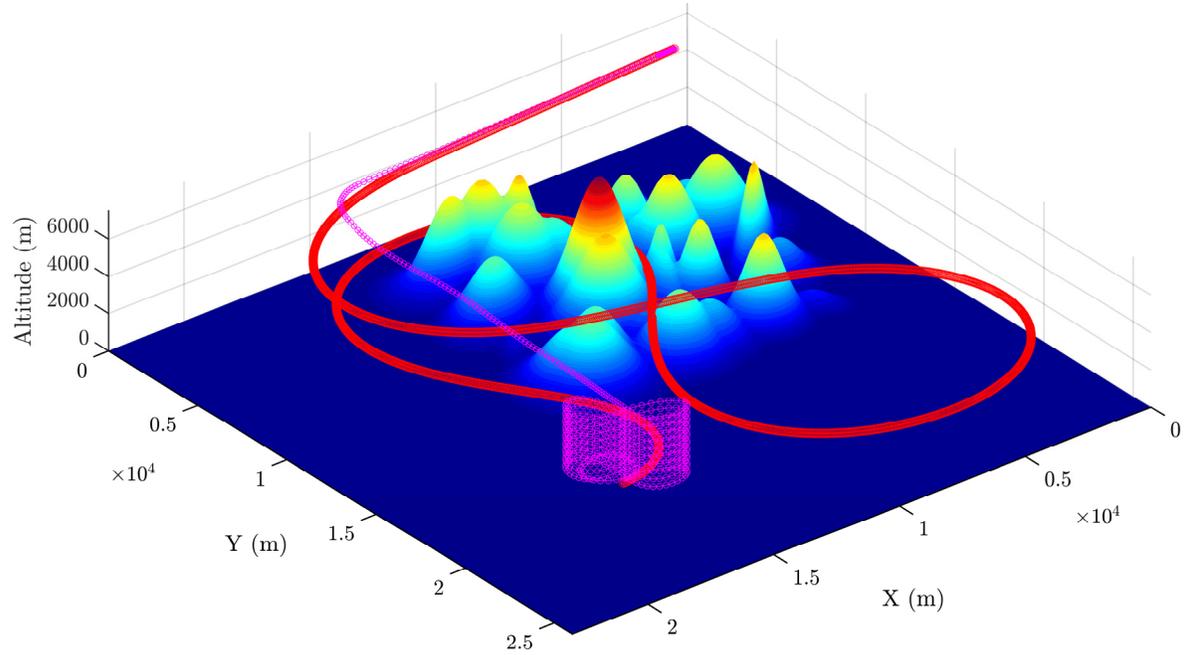

**Fig. 5 3D trajectories generated for impaired GTM with jammed rudder at 10°**

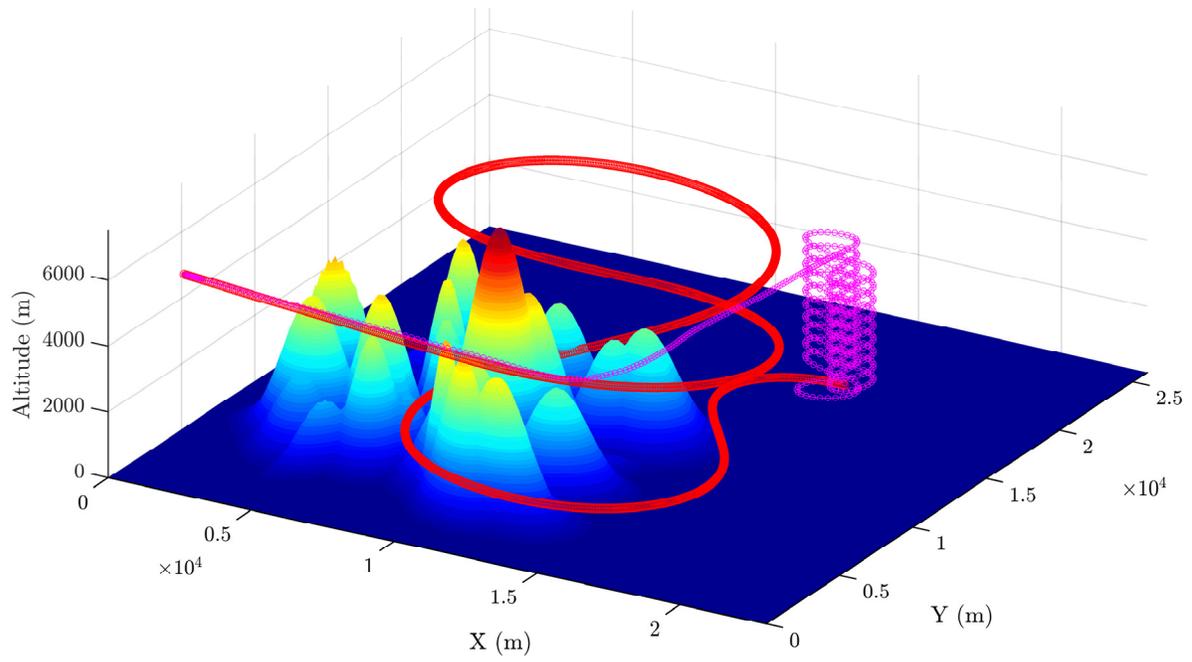

**Fig. 6 3D trajectories generated for impaired GTM with jammed rudder at 10°**

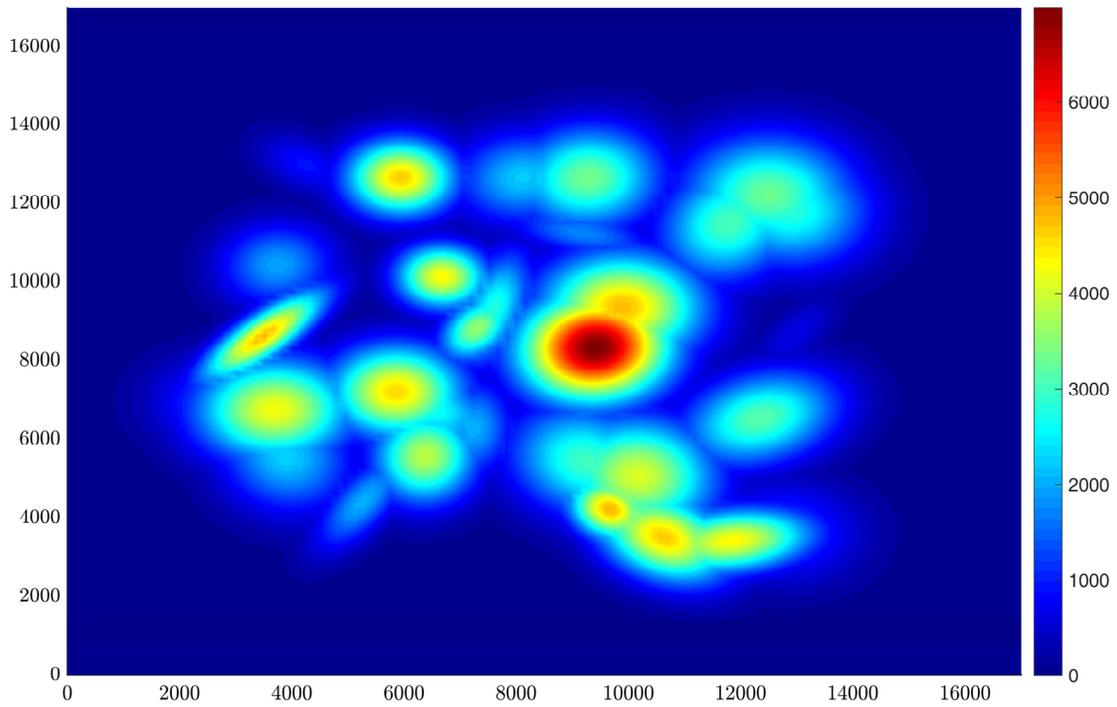

**Fig. 7 Top view of the generated and utilized 3D terrain**

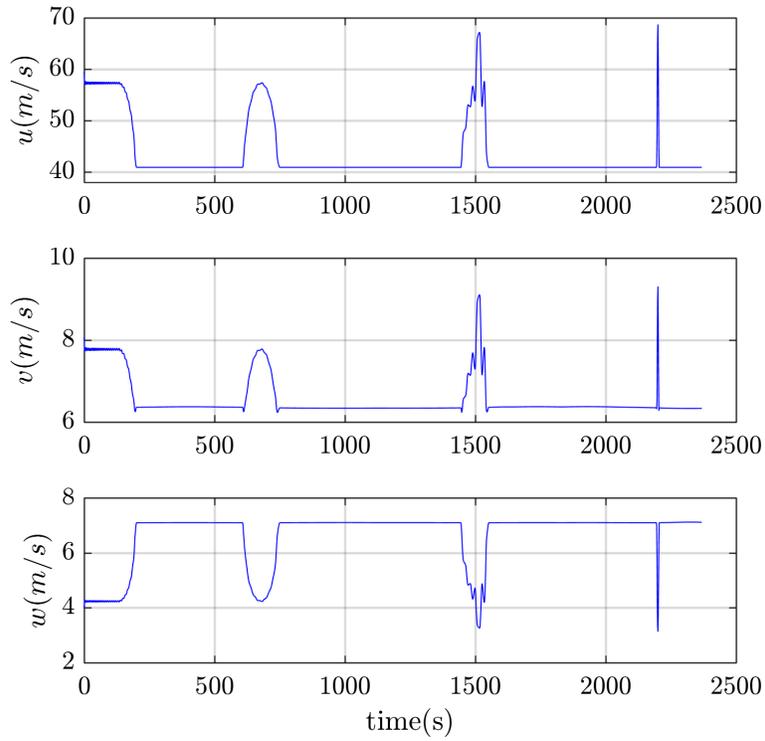

**Fig. 8 Variation of linear velocity components with time (non-real-time trajectory)**

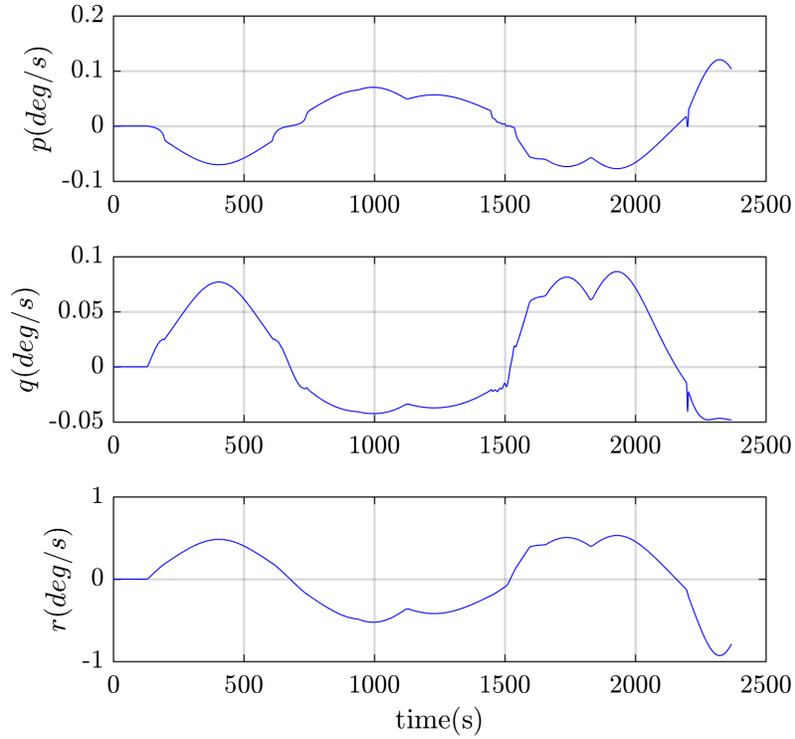

**Fig. 9 Variation of angular velocity components with time (non-real-time trajectory)**

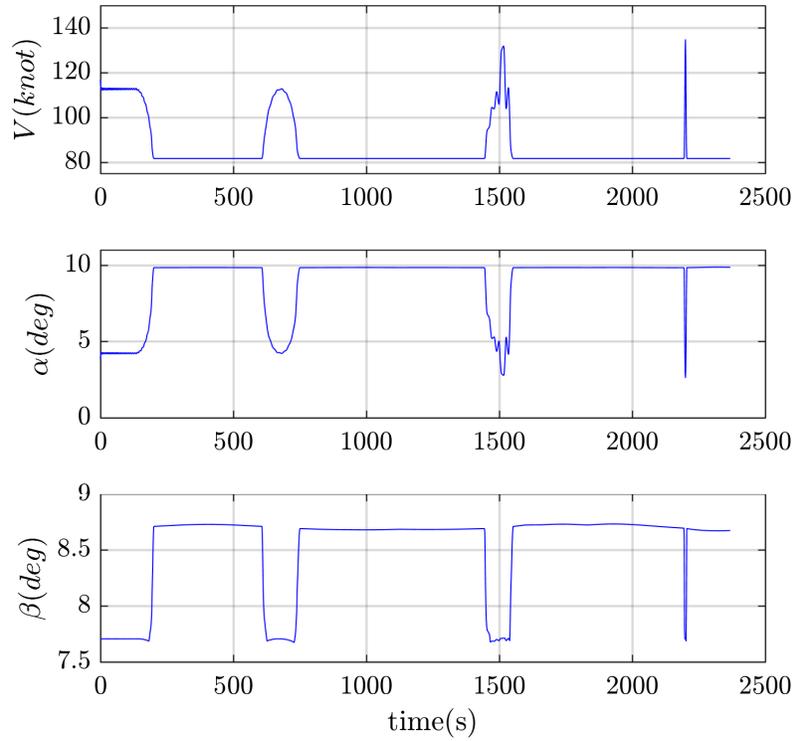

**Fig. 10 Variation of total airspeed, angle of attack and sideslip angle with time (non-real-time trajectory)**

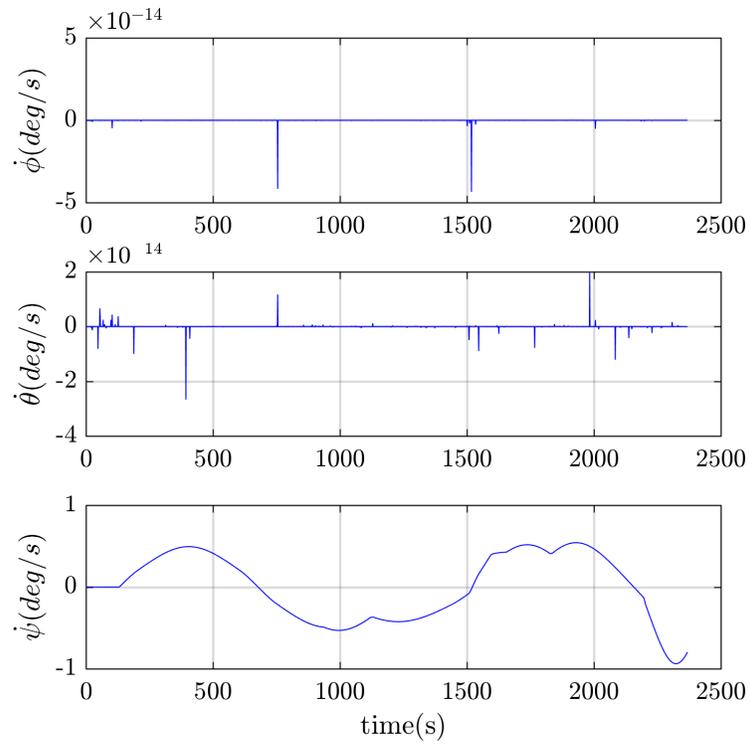

**Fig. 11 Variation of Euler angles rates with time (non-real-time trajectory)**

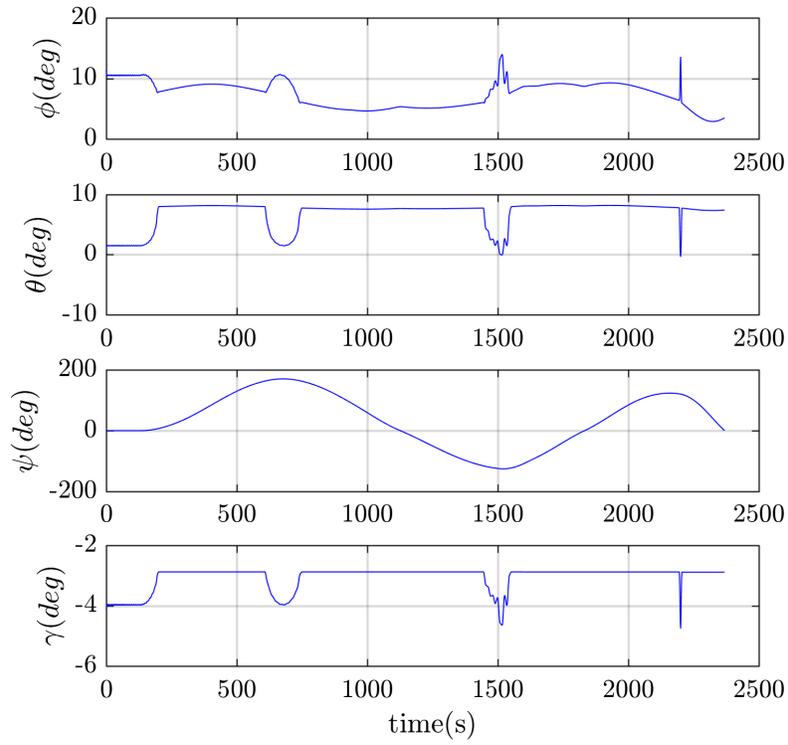

**Fig. 12 Variation of Euler angles and flight path angle with time (non-real-time trajectory)**

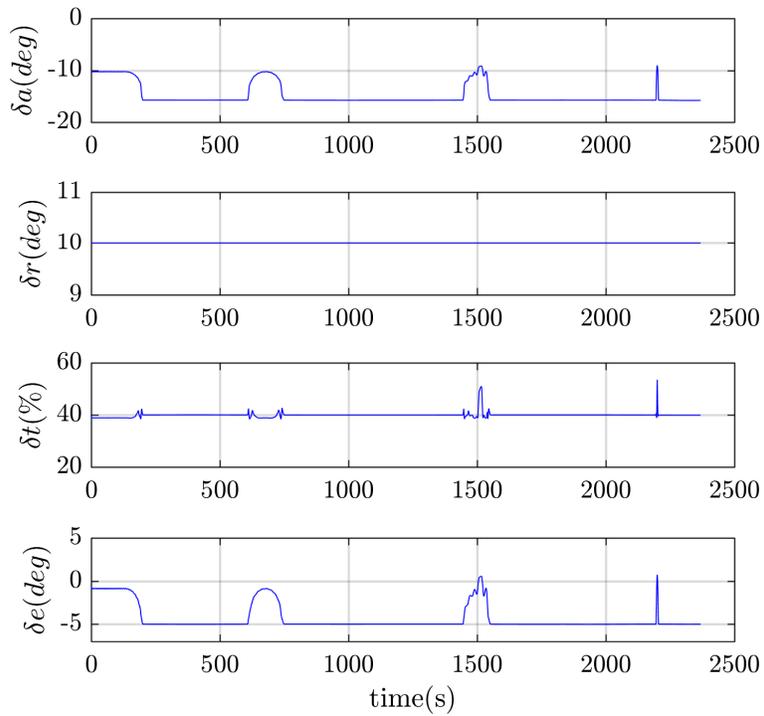

**Fig. 13 Variation of control inputs with time (non-real-time trajectory)**

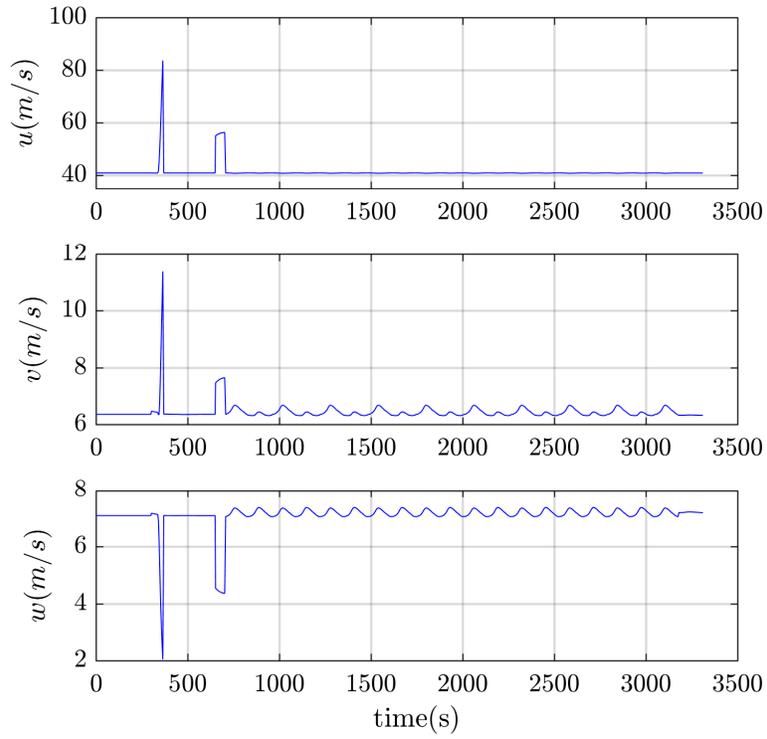

**Fig. 14 Variation of linear velocity components with time (real-time trajectory)**

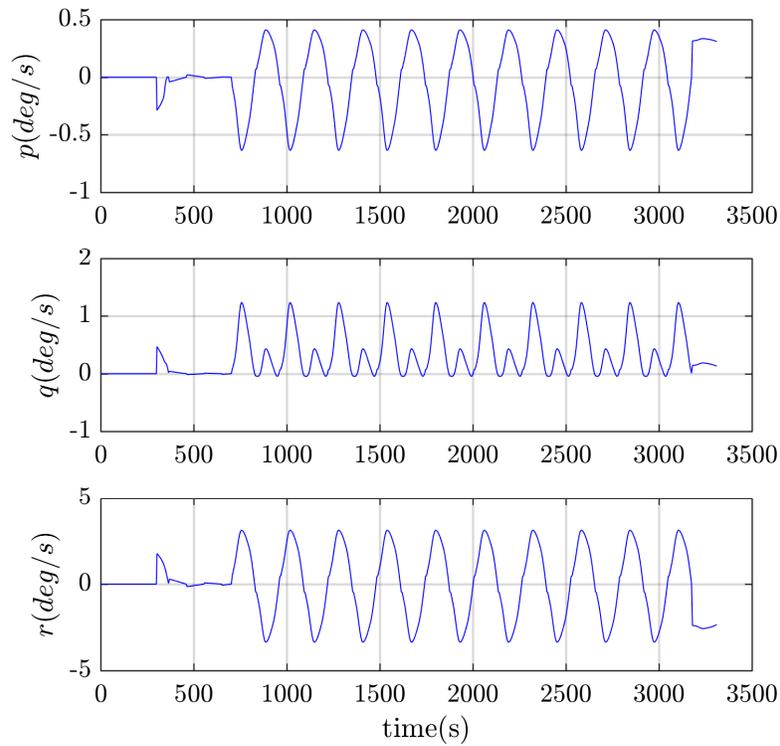

**Fig. 15 Variation of angular velocity components with time (real-time trajectory)**

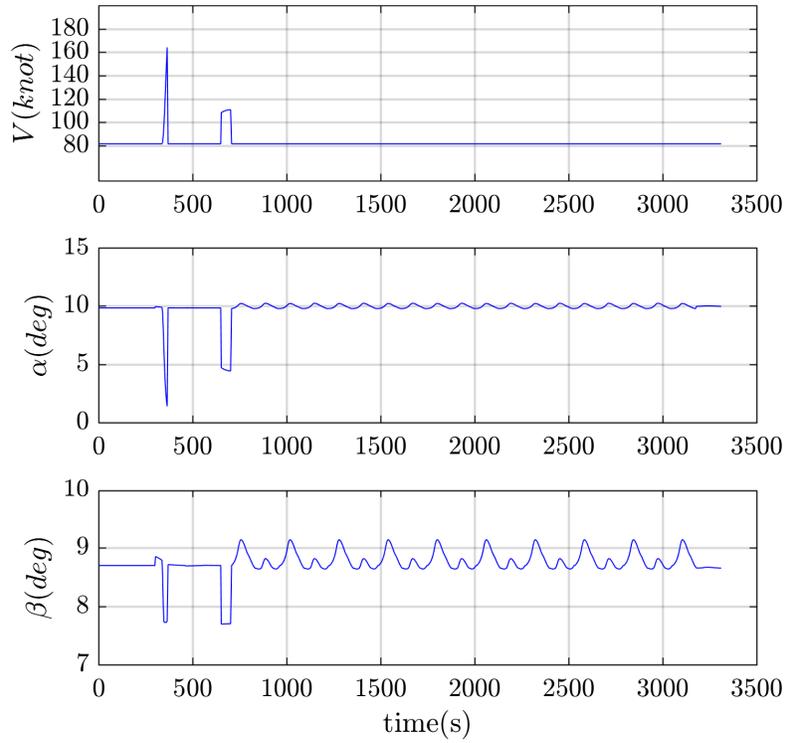

**Fig. 16 Variation of total airspeed, angle of attack and sideslip angle with time (real-time trajectory)**

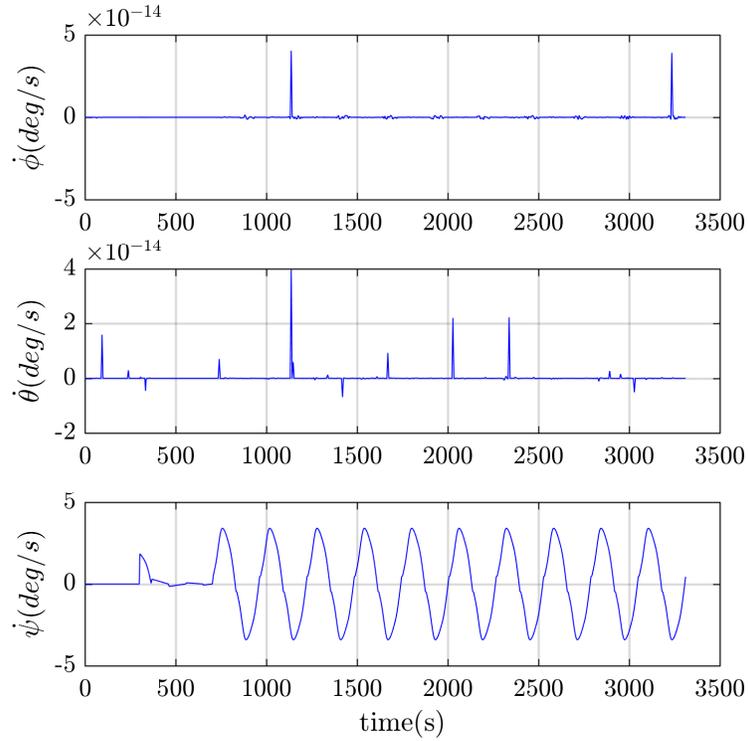

**Fig. 17 Variation of Euler angles rates with time (real-time trajectory)**

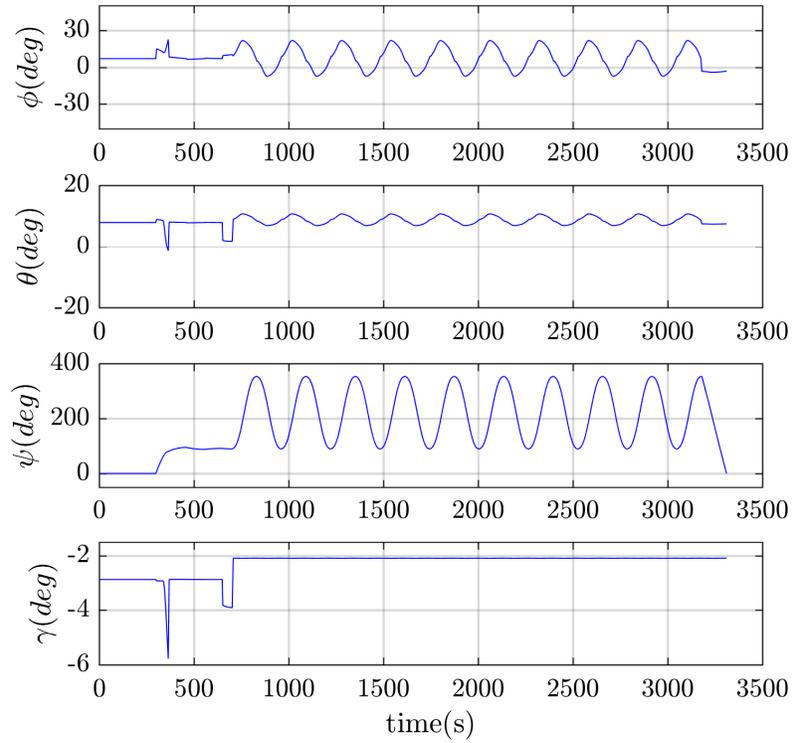

**Fig. 18 Variation of Euler angles and flight path angle with time (real-time trajectory)**

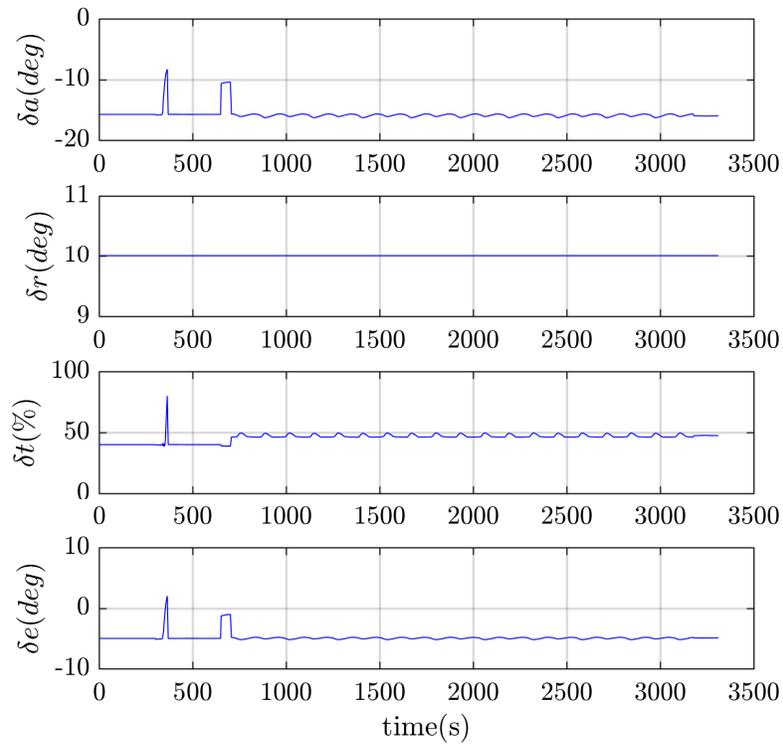

**Fig. 19 Variation of control inputs with time (real-time trajectory)**

The time histories plots (presented in Figures 8 to 19) show that throughout most parts of the path, the aircraft manoeuvres at a speed close to the minimum speed within the manoeuvring flight envelope. Simultaneously, the angle of attack is close to its maximum limit, with the elevator deflection angle set to maintain the aircraft at this angle of attack and the corresponding flight path angle. This approach aims to reduce the cost associated with increasing speed (throttle). Essentially, in most parts of the path the algorithm selects manoeuvres that, considering all conditions, and the coefficients in the cost function, result in a lower final cost. This observation is supported by examining the numerical values of the cost. In fact, based on the considered cost function in the problem and the obtained results, the cost associated with the aircraft flying at a faster speed to generate the required lift—thus flying at a lower angle of attack and subsequently a slightly lower elevator deflection angle—is greater than the cost of reducing speed as much as possible to minimize the throttle contribution, wherein the angle of attack is increased by increasing the elevator deflection angle to compensate for the speed in the lift equation. Obviously, employing other cost functions, such as minimum-time, would yield different results, such as the aircraft descending with the maximum possible speed and flight path angle.

Another noteworthy observation from the results is the temporary increase in speed when the aircraft increases the absolute value of negative flight path angle and descends more rapidly. This happens at approximately 610s, 1445s, and 2195s in the presented time histories of the non-real-time trajectory, and at 340s and 650s for the real-time trajectory. This occurs despite the same manoeuvre (i.e. identical $V, \dot{\psi}$) being attainable in the flight envelope corresponding to a more negative flight path angle (i.e. steeper flight path angle), according to the calculated flight envelopes. In such cases, the speed is slightly increased, while the angle of attack and the elevator deflection angle decrease. This is because, prior to increasing the descent rate, the aircraft is descending with a positive angle of attack (and a positive pitch angle) and a negative flight path angle (i.e. aircraft nose above the horizon), and to transition to a more negative flight path angle, the elevator (which is already deflected upwards to maintain the required angle of attack) is deflected back towards its neutral position, thereby reducing the angle of attack and increasing the absolute value of the negative flight path angle. During this transition, the throttle remains approximately constant. Therefore, the increase in speed is not primarily attributed to compensating for lift (resulting from a decrease in angle of attack). Instead, it is mainly due to a reduction in the weight component along the aircraft's drag force.

It is important to note the significant role of spoilers or speed brakes in the descent of the aircraft. Without the use of spoilers, the aircraft wouldn't be able to descend at high rates because it wouldn't be able to counteract the increase in speed resulting from a decrease (or increase) in the weight component (i.e., with the aircraft nose above or below the horizon) solely by reducing thrust (lowering throttle is not enough by itself to reduce speed sufficiently). However, since the dynamics of spoilers are not well modelled in the GTM model used in this research, they are not included in the equations of motion. As a result, the increase in speed is observed during an increase in descent rate. Table 2 compares the characteristics of the generated paths, the path planning times, the path traversing times, and the final cost in both non-real-time and real-time cases.

**Table 2 Characteristics of the generated trajectories**

| Non Real-time Trajectory | | Real-time Trajectory | |
|---|---|---|---|
| Total Planning Time: | 844 s (14.06 min) | Total Planning Time: | 37 s (0.56 min) |
| Total Traverse Time: | 2368 s (39.46 min) | Total Traverse Time: | 3311 s (55.18 min) |
| Total Nodes: | 400 | Total Nodes: | 150 (5 Nodes per Segment) |
| Total Segments: | 1 | Total Segments: | 30 |
| Total Numerical Cost: | 9594.22 | Total Numerical Cost: | 11985.14 |

As shown in Table 2, the average time taken for path planning for each segment of the real-time trajectory (the piecewise trajectory) is 1.2 seconds. Meanwhile, each segment is traversed in an average time of approximately 110 seconds, as illustrated in Figure 20 depicting optimization and flight time of all 30 segments. Hence, there is ample time available to plan the path for the next segment during the flight of each segment. This is the case even in the shortest segment traverse (52s in this case study). It should be noted that, after selecting the trim points and generating the corresponding three-dimensional trajectory, the control inputs required to execute the selected manoeuvres are calculated within 0.35 seconds for each manoeuvre.

As depicted in Figure 21, the trim points selected by the algorithm for both the real-time (red points) and non-real-time (blue points) paths are inside the manoeuvring flight envelopes of the impaired aircraft.

It is important to consider that the real-time trajectory, aimed at descending the aircraft by 6000 metres, is segmented into 30 segments each with a 200-meter altitude change. The algorithm plans the path for each segment starting from the end of the previous one, gradually reducing the aircraft's altitude by 200 metres with minimum cost while bringing it as close as possible to the runway's starting point on the $x - y$ plane without colliding with terrain. The selection of a 200-meter altitude change per segment is based on an analysis of trajectory optimization results for piecewise paths with different altitude changes per segment. According to the total trajectory optimization cost values presented in the Table 2, the real-time trajectory preserves up to 80% of the optimality. In other words, the cost of the real-time, piecewise, suboptimal trajectory is only 25% higher than that of the one-piece, high-fidelity, optimal path.

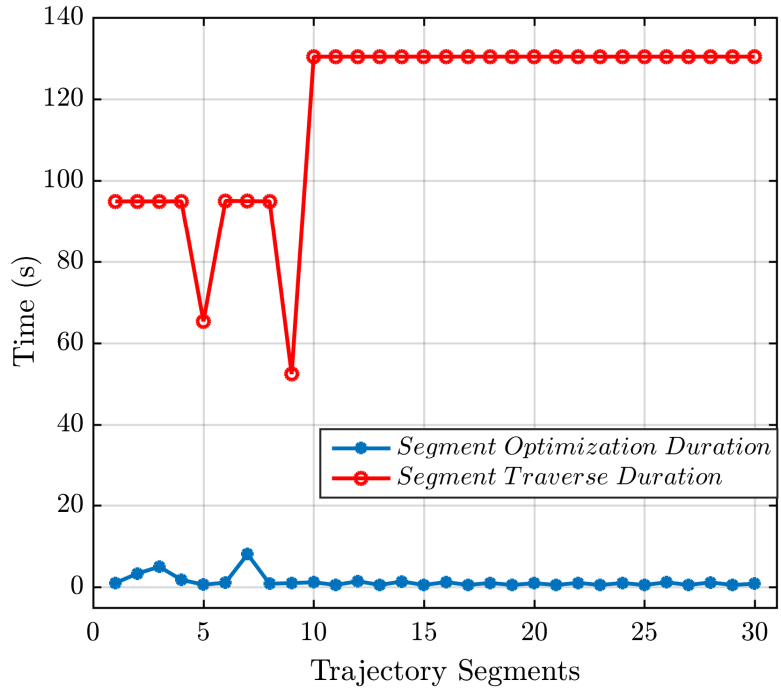

**Fig. 20 Segments traverse and planning times of the piecewise real-time trajectory**

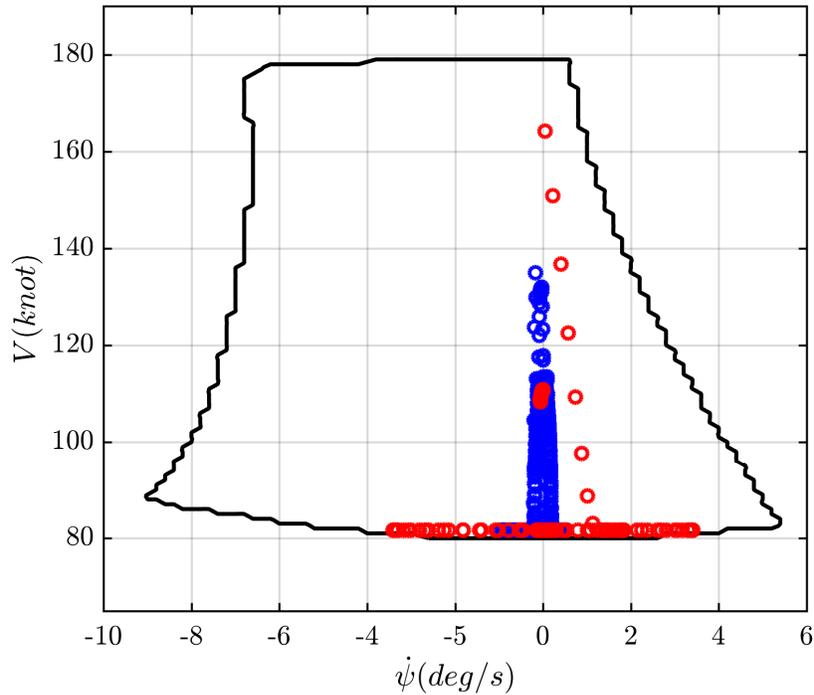

**Fig. 21 Selected trim points by the trajectory optimization algorithm within the impaired aircraft's MFE (the red points indicate the real-time trajectory and the blue points indicate the non-real-time trajectory)**

# IV. Conclusion

In this paper, a method utilizing a pseudospectral method is proposed which enables real-time trajectory optimization for impaired aircraft based on high-fidelity 6-DoF models. The proposed method not only accounts for the reduced manoeuvrability of the impaired aircraft but also takes into account avoiding complex terrain. Specifically, taking advantage of the fact that the manoeuvring flight envelopes (MFE) of the impaired aircraft fall within the boundaries of the unimpaired MFE, a 3-DoF dynamic model is formed with $x, y, z$ and $\psi$ as state variables and $V, \gamma, \dot{\psi}$ as control inputs, enabling high-fidelity path planning for damage cases necessitating non-zero sideslip angle. An optimal control problem is then defined to find trim points within the impaired aircraft's MFE, aimed at optimizing the corresponding generated 3D physical trajectory while ensuring that terrain is avoided. The optimal control problem is solved numerically using the Radau Pseudospectral Method, and 6-DoF dynamic parameters are inversely obtained by trimming the aircraft at the selected trim points of the trajectory utilizing a high-fidelity 6-DoF nonlinear model.

The proposed method was applied to the case of NASA GTM with jammed rudder at $10°$, descending from 6000 metres in a complex mountainous region. A one-piece optimal trajectory was generated, but the computational times exceeded the acceptable range for real-time implementation. Therefore, a piecewise near-optimal trajectory was generated in real time, and the characteristics of both trajectories were compared. Results indicate that for the case study of this research, a terrain-avoiding piecewise real-time trajectory was generated with 25% more cost than the one-piece optimal trajectory.

The manoeuvring flight envelope of an impaired aircraft is contracted due to the imposed failure/damage. A secondary failure which may be induced due to the excessive use of the damaged control surface would shrink the already restricted flight envelope, hence increasing the possibility of loss of control. Therefore, a minimum control effort criterion was adopted for the trajectory optimization. Furthermore, a sensitivity analysis approach was proposed in our previous studies which enables assessing the degree of the effect of different contributing parameters to the variations of the impaired aircraft's manoeuvring flight envelopes and identification of the most effective parameters limiting the impaired aircraft's manoeuvrability. Such a sensitivity analysis method can be incorporated into the trajectory optimization objective such that minimum control effort is imposed on the most influential damaged control surface. This is a potential and interesting topic for future research.

The proposed trajectory optimization approach uses the knowledge of the impaired aircraft's MFE boundaries as path constraints in the defined optimal control problem. This knowledge can be obtained either through the neural network-based real-time MFE estimation method proposed in our previous study or other methods such as an onboard MFE database.

Overall, the proposed novel approach enables near-optimal real-time trajectory optimization for impaired aircraft based on high-fidelity 6-DoF models, whilst considering the remaining manoeuvrability of the aircraft and accounting for complex terrain.

## Funding Sources

This research did not receive any specific grant from funding agencies in the public, commercial, or not-for-profit sectors.